# Data-Driven Methods and AI in Engineering Design: A Systematic Literature Review Focusing on Challenges and Opportunities

Nehal Afifi, Christoph Wittig, Lukas Paehler, Andreas Lindenmann, Kai Wolter, Felix Leitenberger, Melih Dogru, Patric Grauberger, Tobias Düser, Albert Albers, Sven Matthiesen

Karlsruhe Institute of Technology (KIT), IPEK – Institute of Product Engineering
Kaiserstr. 10, 76131 Karlsruhe, Germany

**ORCID IDs**
Nehal Afifi: 0009-0002-7816-1595
Lukas Paehler: 0009-0007-5645-7674
Christoph Wittig: 0009-0008-4124-434X
Andreas Lindenmann: 0000-0002-6566-4559
Kai Wolter: 0009-0002-3613-4476
Felix Leitenberger: 0000-0003-3343-9194
Melih Dogru: 0009-0000-8045-6064
Patric Grauberger: 0000-0002-8367-3889
Tobias Düser: 0009-0005-6652-2818
Albert Albers: 0000-0001-5432-704X
Sven Matthiesen: 0000-0001-5978-694X

**Abstract**
The increasing availability of data and advancements in computational intelligence have accelerated the adoption of data-driven methods (DDMs) in product development. However, their integration into product development remains fragmented. This fragmentation stems from uncertainty, particularly the lack of clarity on what types of DDMs to use and when to employ them across the product development lifecycle. To address this, a necessary first step is to investigate the usage of DDM in engineering design by identifying which methods are being used, at which development stages, and for what application. This paper presents PRISMA systematic literature review. The V-model as a product development framework was adopted and simplified into four stages: system design, system implementation, system integration, and validation. A structured search across Scopus, Web of Science, and IEEE Xplore (2014–2024) retrieved 1,689 records. After screening, 114 publications underwent full-text analysis. Findings show that machine learning (ML) and statistical methods dominate current practice, whereas deep learning (DL), though still less common, exhibits a clear upward trend in adoption. Additionally, supervised learning, clustering, regression analysis, and surrogate modeling are prevalent in design, implementation, and integration system stages but contributions to validation remain limited. Key challenges in existing applications include limited model interpretability, poor cross-stage traceability, and insufficient validation under real-world conditions. Additionally, it highlights key limitations and opportunities such as the need for interpretable hybrid models. This review is a first step toward design-stage guidelines; a follow-up synthesis should map computer science algorithms to engineering design problems and activities.

**Statements and Declarations – Conflict of interest**
The authors have no competing interests to declare that are relevant to the content of this article.

## 1. Introduction

The rapidly growing wave of digitalization is causing a major transformation in engineering design. With the development of connected sensors, industrial IoT systems, and cyber-physical systems, the volume, variety, and velocity of data created throughout the product life cycle are constantly reaching new heights. This “data overload” has caused a paradigm shift in the way various engineering disciplines define problems, draw insights, and develop solutions (Li et al. 2022; Knödler et al. 2023). Simultaneously, the exponential growth and advances in computational power, not only associated with improvements in hardware performance but also with algorithmic advancements in, for example, machine learning (ML), deep learning (DL), and high-dimensional optimization algorithms, now forms the core toolkit of modern engineering analytics (Bach et al. 2017b). The multitude of available data combined with powerful and affordable infrastructure for data processing facilitates the widespread adoption of data-driven methods (DDMs) from conceptual design and requirements engineering to validation and optimization (Shabestari et al. 2019; Gay et al. 2021). Despite their current widespread adoption, a high diversity in understanding and uncertainty in application of DDMs remains. In the following, we summarize the understanding of DDMs and the alignment of the application of DDMs to the V-Model (VDI/VDE 2206, 2021) as basis for deriving the research gap.

## 1.1 What are DDMs?

In Literature, DDM are defined in several distinct ways, from leveraging empirical data (Maslyaev et al. 2020; Talal et al. 2020) to the process of data-driven innovating drawn from big data (Luo 2023), combination of data analytics techniques (Ayensa-Jiménez et al. 2019) or information for and enhancement of decision making, modelling and analysis (Nosck et al. 2023). More broad definitions describe DDMs as methods to extract knowledge from data to support modeling, decision-making, prediction and optimization (Wu 2024; Payrebrune et al. 2024). Furthermore, (Knödler et al. 2023) frame them as analytical systems capable of adapting dynamically to incoming data. Another line of research defines DDMs by how they learn from data. These studies infer empirical relationships from large datasets (Mosallam et al. 2015; Villarejo et al. 2016) using machine learning and related methods in computational intelligence algorithms (Mount et al. 2016). (Gao et al. 2013)) argues that DDMs can operate with limited prior process knowledge when they rely on, for example, signal processing and large-scale analytics. Building on this view, researchers emphasize reduced dependency on domain specific background knowledge when prior theoretical knowledge is limited and system complexity is high (Mount et al. 2016; How et al. 2019). This orientation is consistent with software engineering view that distinguish data driven Learners, that infer behavior from data, slow to train, and fast to run, and model-based Solvers, that compute with explicit models, quick to start, and slower to run (Geffner 2018). So, once DDMs are trained, they may produce outputs faster than methods that depend on explicit models (Geffner 2018). Furthermore, a widely cited and general definition, originating from ML research by (Jordan and Mitchell 2015) and adopted in this publication, describes DDMs as methods to derive insights or control actions directly from data, without relying on traditional engineering models. Traditional engineering models can be, for example, based on physical knowledge and assumptions, as well as analytical formulations. In summary this means that there is no standardized definition of DDMs in engineering design. But it also reflects the growing recognition of DDMs as not only tools to handle big data, but critical enablers of modern, adaptive, and scalable engineering design solutions.

## 1.2 Alignment of the application of DDMs to engineering design stages using the V-model

Within engineering design, there is a variety of different models that describe its processes and stages (Wynn and Clarkson 2018). One of the widely recognized process models is the so-called V-model. The initial idea of the V-model was first introduced in software development by (Boehm 1979) and later taken up by (Bröhl and Dröschel 1995). It provides a systematic guideline for product development, encompassing various stages from requirements analysis to verification and validation. For mechanical and mechatronic systems, this requires the integration of mechanical, electronic, and software components, necessitating a multidisciplinary approach. This leads to explications like the VDI 2206 guideline - a V-model-based standard in mechatronic and cyber-physical systems development considering different domains of engineering design shown in Figure 1 (VDI/VDE 2206, 2021). Recently, development frameworks such as the Double-V, Triple-Vs model (Li et al. 2019b) and the AI4PD ontology (Gerschütz et al. 2023a) have emerged to systematize the integration of data, models, and product development lifecycles. The transition from the traditional V-model to the Double and Triple-Vs models reflects a growing need to synchronize product lifecycle stages with real-time analytics and adaptive behavior toward the use of DDMs. In this publication, we summarized the product-development stages to system design, system implementation, system integration, and validation to investigate different types of applications of DDMs.

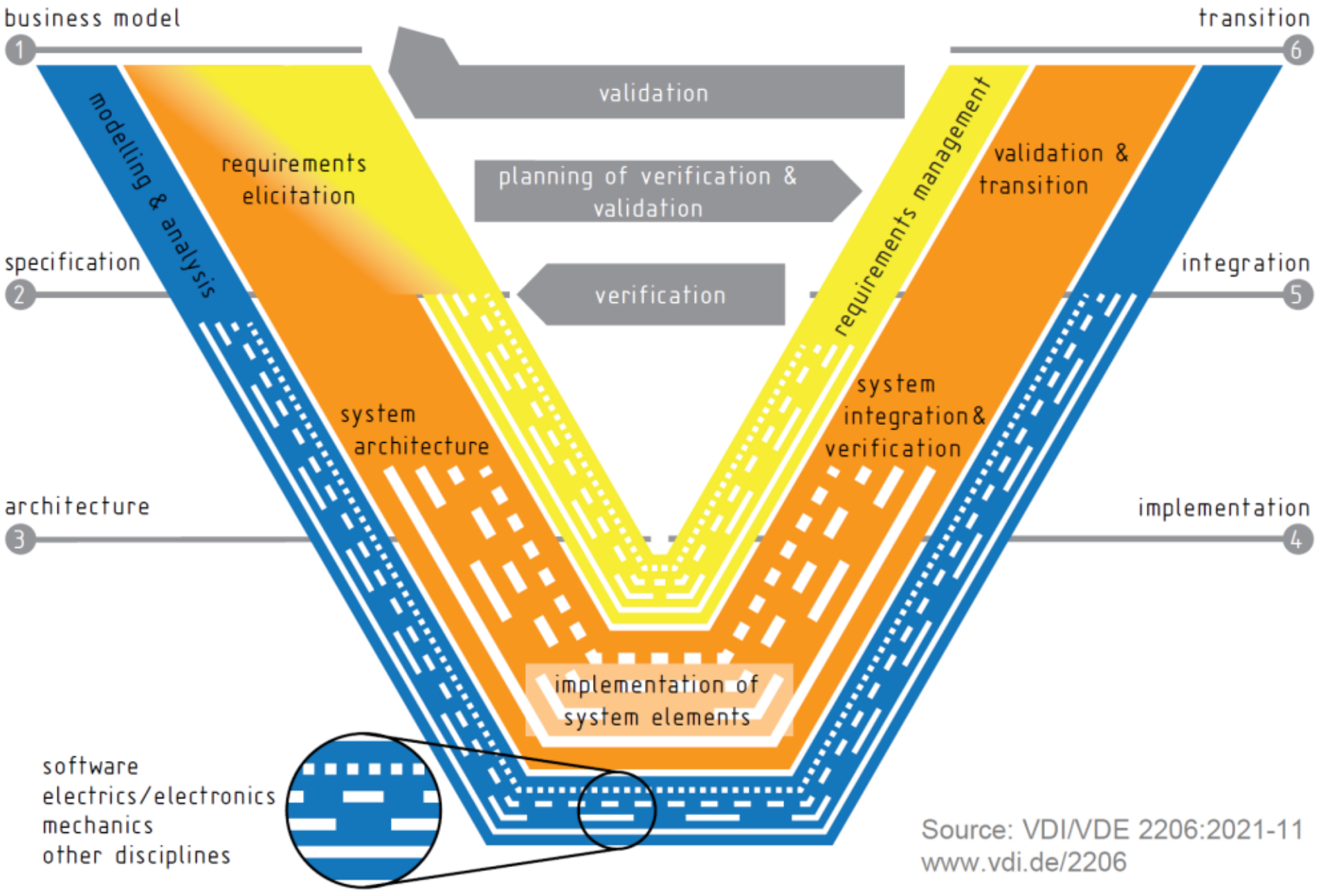

Figure 1 An overview of the V-model for product development of cyber physical systems (CPS) guidelines based on (VDI/VDE 2206, 2021)

## 1.3 Research Gap

DDMs have already proven potential in engineering design; for instance by enabling performance prediction, anomaly detection and design optimization across early and late product-development stages (Li et al. 2019b; Jacobs et al. 2022). Hybrid approaches that combine empirical data with expert or physics knowledge are also gaining popularity, one example in prognostics and health management (Gay et al. 2021). Likewise, (Verma and Singhal 2024) emphasize the role of Artificial Intelligence (AI) in enabling adaptive decision-making in complex systems. Yet adoption remains uneven, based on (Shabestari et al. 2019)) most applications of DDMs are confined to requirements modelling or conceptual design, while downstream activities such as prototyping, system integration and validation receive far less attention. This imbalance indicates that the engineering design community still lacks clear, product development life-cycle-wide guidance on how, when and with what evidence to deploy DDMs. Recent literatures tried to address this. (Figoli et al. 2025) presented two decades of “AI-in-design” literature through a bibliometric map, clarifying who researches what, yet the study offered no link between thematic clusters and concrete engineering tasks. On the other hand, (Payrebrune et al. 2024) combine the V-model for product development with AI tools and showcase eight dynamical-system cases; their insights are rich but rest on a narrow sample from a defined technical domain which limits generalizability. Furthermore (Nüßgen et al. 2024) propose robustness criteria for context-sensitive AI within the V-model; however, their review covers few numbers of studies and does not provide any quantitative assessment of method maturity. (Fang et al. 2025) focus exclusively on generative AI in conceptual design, offering valuable insights for that one phase while leaving later stages like integration, and validation untouched. (Gerschütz et al. 2023b) look broadly at digital-engineering use-cases across V-model stages. Their survey offered a useful head-count of ML applications along the V-model stages and showed that most published work concentrates on the left-hand branch; system design and implementation. Helpful as that finding is, the paper does not classify the DDMs, or say which algorithms were actually used, or what data types were fed to them, or what limitations or opportunities each approach has. These studies try to examine how data transitions across consecutive design stages, yet there is still lack of a holistic perspective.

Collectively, these reviews confirm and highlight the importance of DDMs in engineering design yet still fall short in delivering a stage-specific, quantitative map that aligns specific DDMs, algorithms, data type, application, limitation, and opportunities across the mechanical and mechatronic product development.

The problem still lies in the lack of clarity on "what" and "when" to use DDMs in product development which arises from the growing variety of these methods, hindering their practical application. But to develop such a guideline, an overview of the current state of research is needed as a first step. Without such a synthesis, researcher and engineering designers must still rely on implicit, experience-based choices rather than research- and evidence-based guidance when selecting DDMs during different product development stages. Accordingly, there is need to create a comprehensive, product development-aligned synthesis that links concrete DDMs, their current data requirements, application areas, and systemic challenges of each stage, thereby collect scattered insights into initial guidance.

## 1.4 Contribution

To address this problem, this article conducts a systematic literature review regarding the application of DDMs in mechanical and mechatronic product development. It captures this across V-model for product development activities from early design to validation. The study is guided by the following overarching research question (RQ):

***RQ:*** *What is the current state of application, challenges, and opportunities in applying DDMs across the product development lifecycle of mechanical and mechatronic systems?*

This research question is investigated through the following three sub-questions (SQ):

- **SQ 1:** How are DDMs implemented in engineering product development, and what data types and algorithms do they rely on?
- **SQ 2:** How are these DDMs applied across the stages of the product development process?
- **SQ 3:** What challenges limit, and what opportunities enable, the effective use of DDMs across the product development lifecycle**?**

By answering these questions, the study provides a structured overview of current DDM use, identifies areas of limited adoption, and highlights challenges and opportunities. It offers the initial step for having guidance to support engineering designers and researcher on where and when DDMs can be effectively applied in product development. The remainder of the article is structured as follows: Section 2 outlines the review design, including the stage classification framework and selection criteria. Section 3 presents the results, structured around both cross-stage and stage-specific analyses. Section 4 discusses the findings in relation to the research questions and identifies implications for research and practice. Finally, Section 5 concludes key insights and future directions.

# 2. Methodology

To systematically investigate how DDMs are integrated into the product development of mechanical and mechatronic systems, a structured literature review was conducted following the PRISMA 2020 guidelines (Page et al. 2021), incorporating both quantitative and qualitative synthesis. This approach supports transparent and reproducible identification, selection, and analysis of relevant literature. To further guide stage-wise classification and analysis, the V-model (VDI/VDE 2206, 2021) was adopted.

## 2.1 V-Model Stages

For simplification, this review focuses on four core development stages, illustrated in Figure 2, namely: system design, system implementation, system integration and validation based on (Gerschütz et al. 2023b). Each stage is associated with a set of characteristic activities, which formed the conceptual basis for the stage-specific search terms used in the review (VDI/VDE 2206, 2021). In ***System design*** an overall solution structure of the system is developed based on the specified requirements, which includes the interaction between the requirements, functions, logical and physical structure. This results in implementable units of the system to which requirements and interface information are assigned. It contains activities that contribute to the decomposition of the system into implementable units. These activities include the comparison of alternative solution approaches with regard to various criteria, the inclusion of users in the consideration of solution approaches and the decomposition of the system from a functional, property and implementation perspective.

In ***System implementation***, translates the general system design into detailed design components and subsystem suitable for manufacturing, integration, and testing in the subsequent steps of product development. It includes activities like, but are not limited to, geometric computer-aided design (CAD), programming, dimensioning, tolerancing, design optimization, and simulations such as Finite-Element Analysis and Multi-Body Simulations. The outcome is a functional physical product or prototype. In ***System integration***, previously developed subsystems, modules, and components are assembled into a functioning system, and their interactions are evaluated to ensure compliance with system-level requirements. It focuses on verifying interfaces, managing interdependencies, coordinating integration activities, verification, and testing system behavior under combined operational conditions. Both subsystem performance and collective system behavior are assessed, often under complex operational and boundary conditions. ***Validation*** assesses whether it meets stakeholder requirements and fulfills its function. Unlike verification, which addresses specification compliance, validation evaluates real-world performance. This phase represents the culmination of the development process on the right side of the V-Model, where the system's functional behavior is comprehensively assessed. Key activities include system tests to assess the complete system, field tests to examine real-world behavior, and acceptance tests to confirm the fulfillment of agreed-upon criteria. Additional activities include performance evaluations, reliability assessments, and user studies. This V-model adoption enabled classification of studies and informed the search strategies.

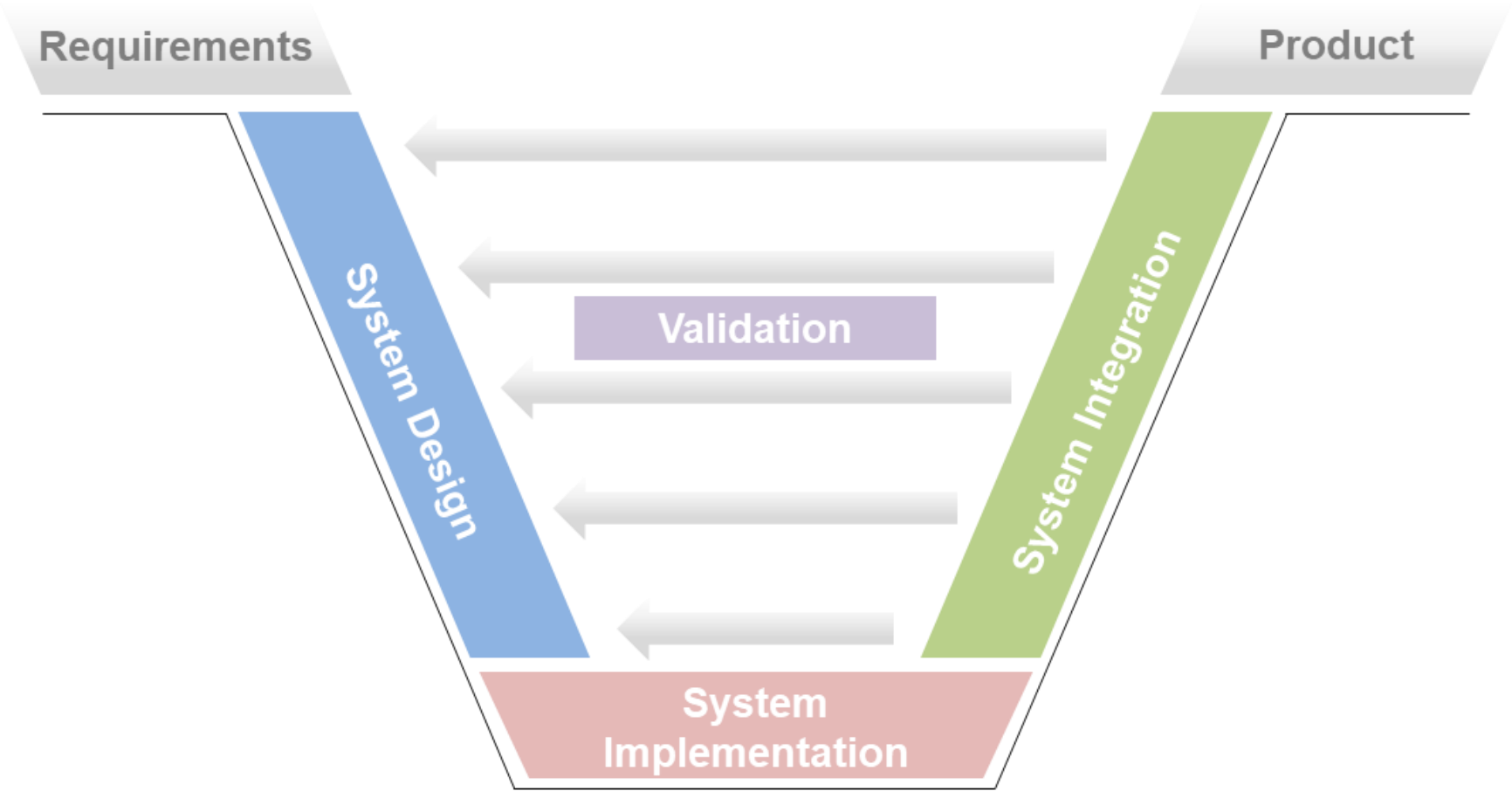

Figure 2 Illustration of the simplified V-model for product development (Gerschütz et al. 2023b)

## 2.2 Review Design and Data Sources

The review process was divided into four main phases: (1) database selection and search string formulation, (2) literature retrieval and de-duplication, (3) eligibility screening, and (4) full-text coding and synthesis. First, a multi-stage search strategy was conducted across the three major academic databases: Scopus, Web of Science, and IEEE Xplore. The search aimed to identify articles published in English from January 1$^{st}$, 2014 to July 1st, 2024 and limited to the engineering domain. The final search was executed on July 1st, 2024. To support classification and guide search string formulation, each stage of the was operationalized by identifying its core activities based on the (VDI/VDE 2206, 2021). A core search string was developed to capture publications relevant to DDMs and product development as seen in Table 1. ML, DL, and statistical/probabilistic methods were selected as the three main classifications of data-driven methods for this review. As illustrated in Figure 3, ML is a subset of AI and DL is a further subset within ML, reflecting increasing specialization and complexity. In parallel, statistical and probabilistic methods intersect with all levels, forming the theoretical foundation upon which modern AI approaches are built. These terms reflect methodologies that are foundational to AI but are more precise in describing algorithmic approaches used in engineering contexts. Their inclusion ensures that the search strategy targets technically grounded contributions, rather than generic or conceptual references to AI. This core string was extended with product development stage-specific strings, tailored for the four primary V-model product development stages as seen in Table 1. After retrieval, intra-stage duplicates were removed first, followed by inter-stage de-duplication to account for studies that span multiple development stages. Cross-stage duplication was handled by assigning publications to all relevant stages without inflating the total count.

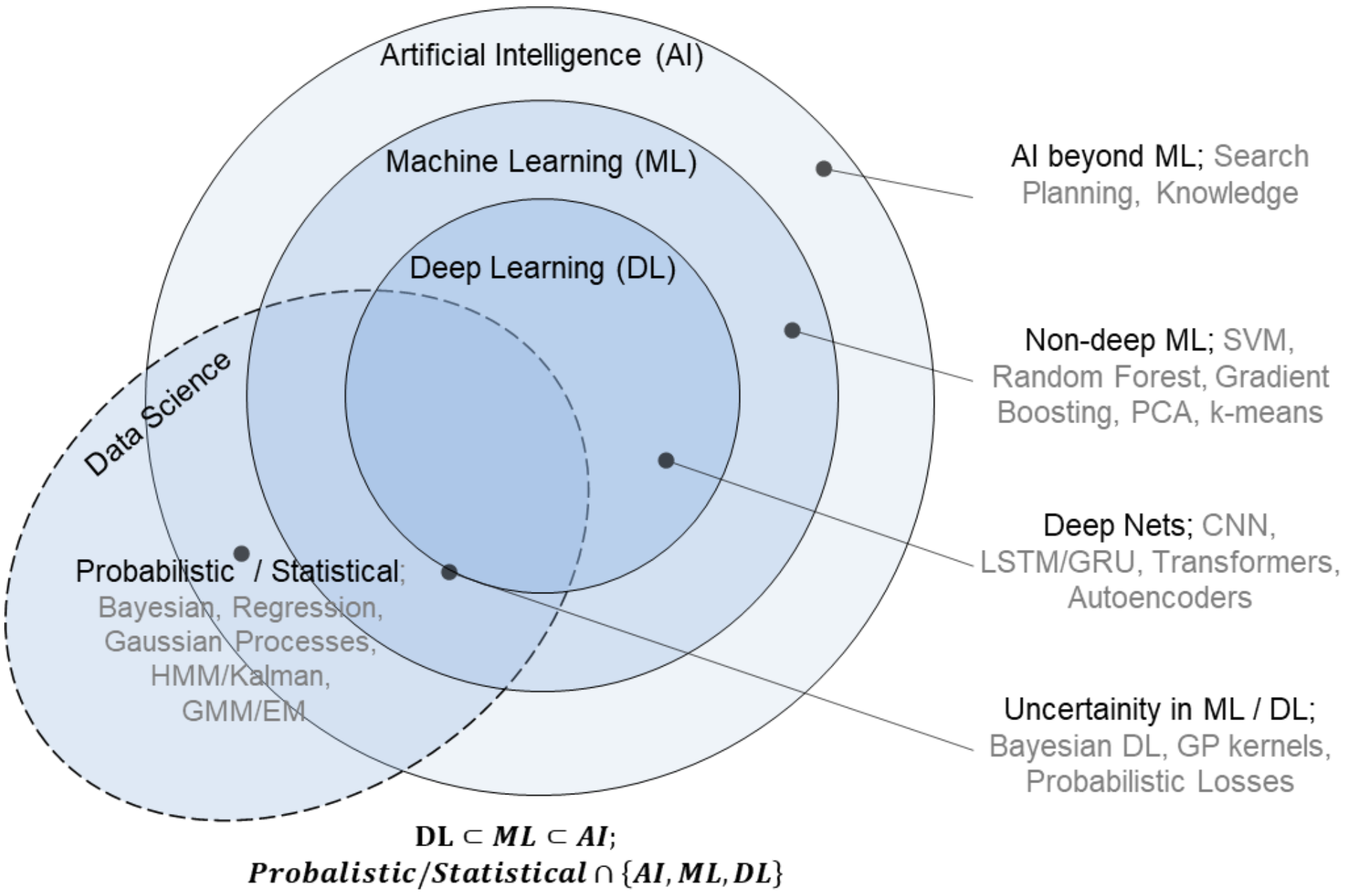

Figure 3: Classification of DDMs highlighting the nested structure of AL. ML and DL, and the foundational role of probabilistic/statistical approaches. Used to support the selection of search terms.

Table 1 The Search String used to extract research articles from the three databases in which each team uses the core search string plus their respective team search string

| ***Area*** | **Search String** |
|---|---|
| ***Core*** | (data driven \| data-driven \| statistical \| machine learning \| data mining \| deep learning) **AND** (product development \| product engineering \|mechanical system* \| mechatronic system \| system* engineering |
| ***System Design*** | **AND** (system design \| system specification \| requirement \| requirements \| concept) |
| ***System Implementation*** | **AND** (system implementation \| FEM \| finite element method \| CAD \| computer aided design \| tolerance design \| tolerancing \| design optimization \| system simulation \| dimensioning) |
| ***System Integration*** | **AND** (integration) **AND** (test* \| predictive* \| fault \| analytics \| verification \| reliability \| system integration method) |
| ***Validation*** | **AND** (validation \| verification) |

## 2.3 Inclusion and Exclusion Criteria

Furthermore, after removing duplicates, a two-stage screening process was applied (Page et al. 2021), the. In the first screening stage, titles, abstracts, and keywords were reviewed using the following inclusion criteria: domain relevance to mechanical or mechatronic product development, mention of DDMs or specific algorithms, application in product development activities, and focus on the product development rather than pure algorithmic development. Furthermore, exclusion criteria were set to papers related to electric systems, energy and engineering management. In the second screening stage, full texts were evaluated for alignment with the overarching research question and whether the key activity in the article is mapped to each respective product development stage based on the explained operational criteria of the (VDI/VDE 2206, 2021).

## 2.4 Data Extraction for Quantitative and Qualitative Analysis

To further evaluate the included articles, a structured data extraction procedure was employed to enable both quantitative and qualitative analysis. Each publication was analyzed according to predefined inclusion and exclusion criteria. First, the type of DDMs was identified and categorized as machine learning, deep learning, statistical, as illustrated in Figure 3, or hybrid approaches that uses mix of ML or DL with statistical method or methods that incorporate the mathematical element such as physics informed neural networks (PINNs). Second, the algorithmic technique used within each study was recorded. Third, the type of input data was extracted. Finally, the engineering application domain or task was examined. These extracted dimensions served as the basis for stage-wise classification and facilitated the identification of usage trends, methodological patterns, and gaps in the current application of DDMs across the product development lifecycle, thereby supporting the analysis required to address the research questions. Additionally, a qualitative assessment of the 114 included articles was conducted to evaluate their methodological transparency.

The evaluation followed the trustworthiness framework proposed by (Guba 1981), extended by (Guba and Lincoln 1982), and fully elaborated in (Lincoln and Guba 1985) this evaluation focused on four dimensions to assess the quality of research across disciplines: credibility, transferability, dependability, and confirmability. The fourth dimension, confirmability, which addresses researcher neutrality, was excluded as it is not directly applicable when analyzing secondary literature. Extending this approach, ethical considerations were introduced as a distinct criterion to reflect current expectations around research integrity, especially in applied and interdisciplinary engineering contexts. Although ethics were only indirectly addressed in the original framework by (Lincoln and Guba 1985) under reflexivity, appraisal tools such as COREQ (Tong et al. 2007) and CASP (Long et al. 2020) now explicitly incorporate ethical aspects such as research approval, consent, and responsible data use. Including this dimension allows for a more comprehensive assessment of methodological rigor. Furthermore, credibility was assessed by checking whether studies described how their results were validated or verified. Transferability examined whether sufficient contextual detail was provided to enable replication or application in other settings. Dependability referred to the transparency of the research process and whether methodological choices were clearly documented. Ethical considerations were assessed by determining whether studies reported research ethics approval, data consent, or responsible data use. To complement the qualitative findings, we extracted the number of references and citation count for each included article. This allowed us to evaluate the extent to which each study engaged with existing literature and to observe trends in scientific visibility. While these metrics do not directly reflect methodological quality, they offer complementary insights into how the work is positioned and received within the field.

# 3. Results: Stage-wise Analysis

First, an overview of the detailed results regarding the four stages are presented in Figure 4. The database search (2014–2024) returned 1689 records; after duplicate removal, first screening and second screening 114 studies were retained for full analysis, 10 of them are appeared and analyzed in two different stages. Figure 4 summarizes the outcome of the screening process and the allocation of the 114 eligible studies to the four simplified V-model stages. The distributed as follows: 52 papers address system design, 30 focus on implementation, 26 on integration and 20 on validation. The last stage in Figure 4 illustrates the number of articles between stages. In which three articles appeared in system design and implementation, four in system design and integration, one system implementation and integration and two in system integration and validation Figure 5 shows the temporal distribution of articles applying DDMs in product development. A clear upward trend is observed over the decade, with an increase beginning in 2017 and peaking in 2021.

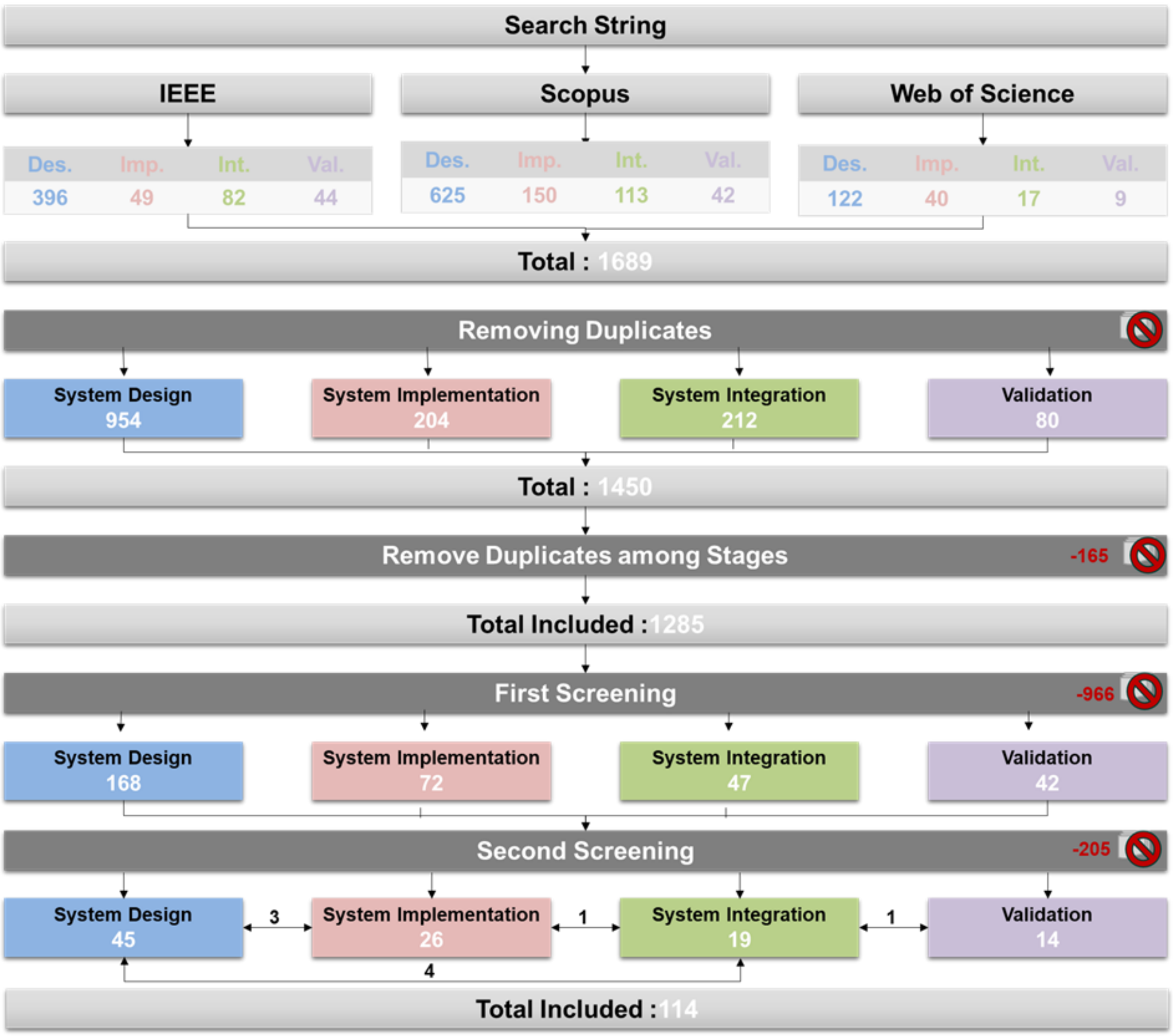

Figure 4: Overview of the Systematic Review results for product development stages regarding each step

## 3.1 Qualitative Analysis

The results of the qualitative assessment and overview are described in in Figure 6, and 7. Figure 6 summarizes the qualitative results with respect to each stage. A clear variation was observed across the reviewed studies in terms of how well each criterion was addressed. While credibility was frequently reported, often through the article validation methods or detailed methodology sections. Dependability and transferability, on the other hand, were less consistently covered fall off by roughly 20 %, with many studies lacking sufficient contextual detail to support generalization. Notably, explicit discussion of ethical considerations is rare or hard to assess, marked “unclear” in almost all cases. Only a minority of studies explicitly mentioned ethics approval, consent, or data governance practices, only in cases where human-centered methods or user evaluations were involved. Furthermore, as described in the methodology, the number of references and citation counts were extracted to provide additional context on literature engagement and visibility.

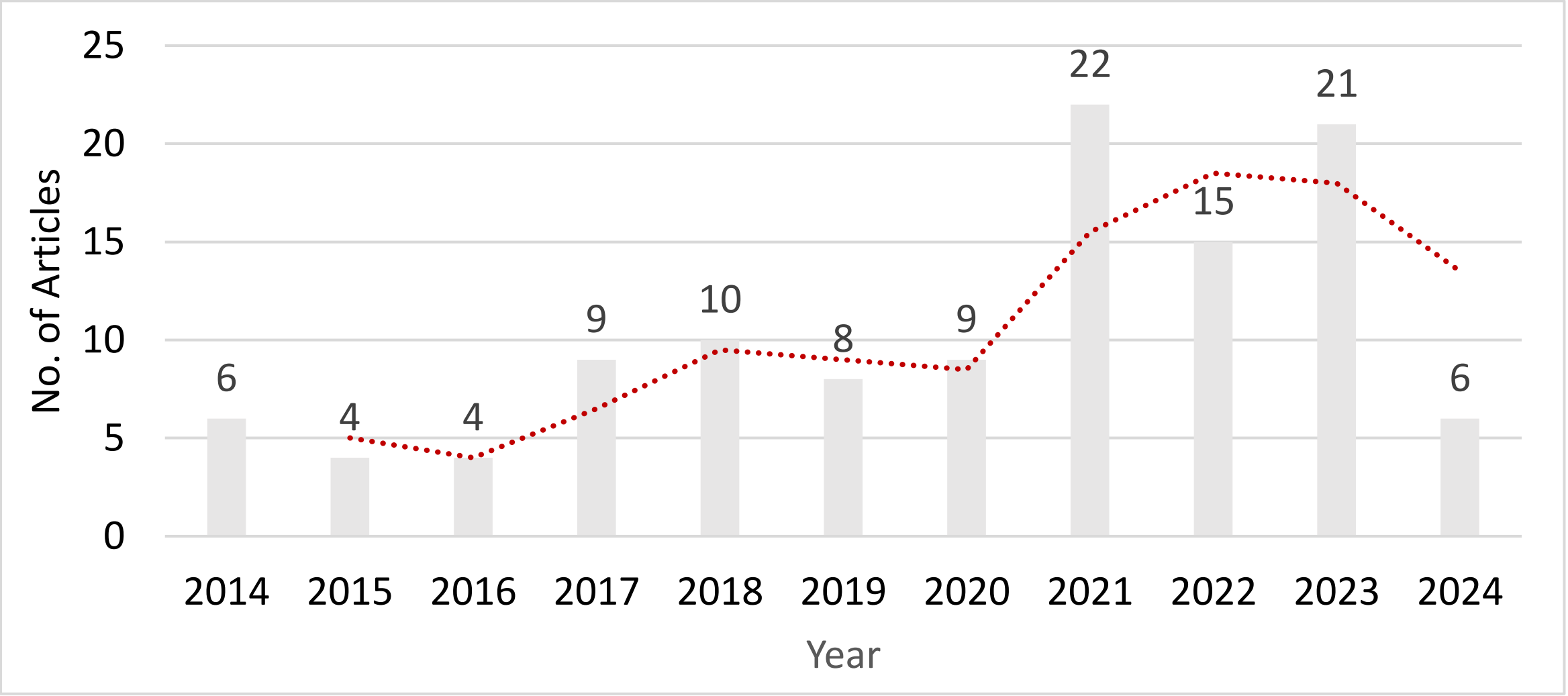

Figure 5: Yearly distribution of included articles (2014–2024) with a trendline showing growth peaking in 2021–2023.

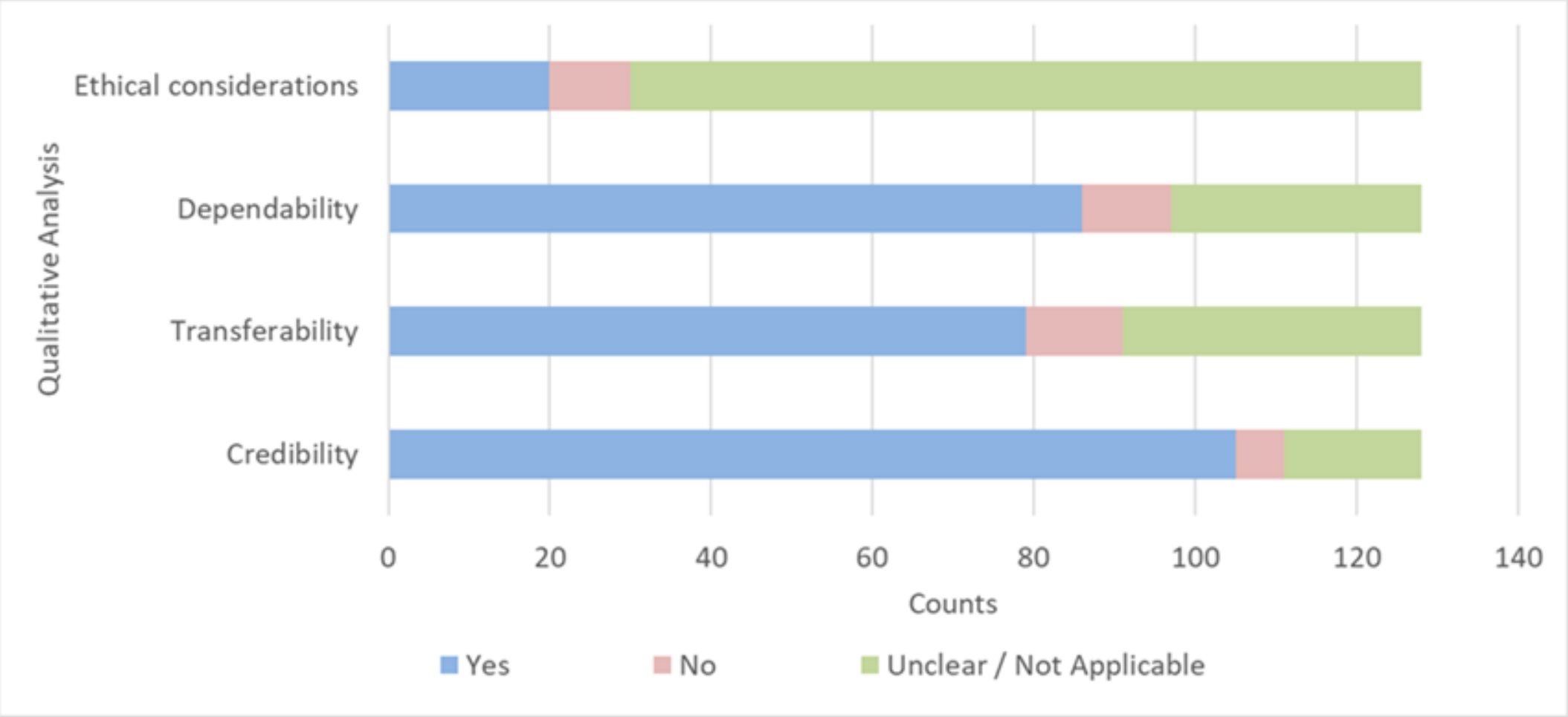

Figure 6: Results of the qualitative analysis of the included articles.

Figure 6 groups the papers by their reference list length and citation count; most studies cite 21–60 sources but have fewer than 20 external citations. This is consistent with the publication timeline shown in Figure 5, where a substantial portion of the reviewed studies were published in 2021, 2022 and 2023. The lower number of articles shown in 2024 likely reflects an incomplete publication year rather than a reversal of the upward trend. As citation accumulation typically lags behind publication, especially in engineering, the lower citation counts observed are likely influenced by the recency of many articles in the dataset.

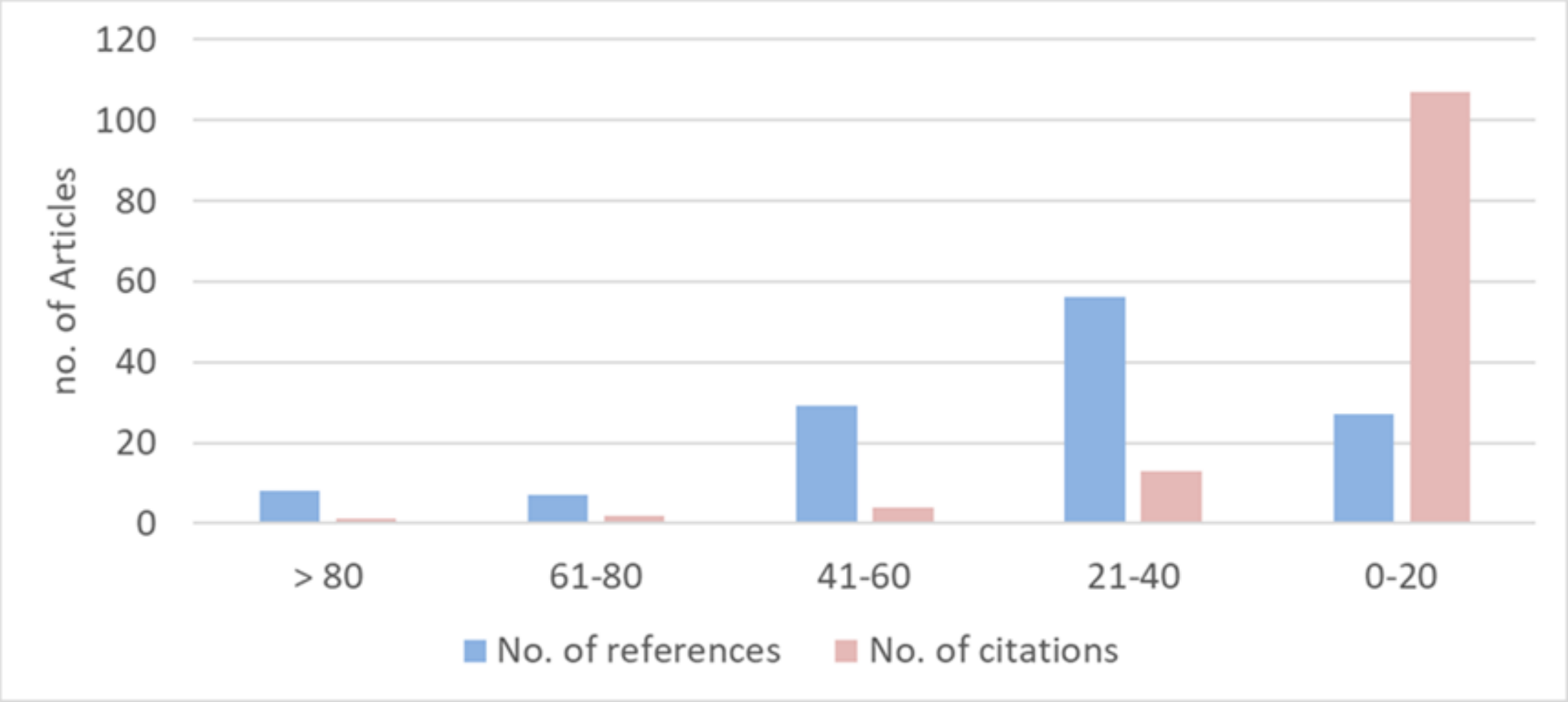

Figure 7: Distribution of the number of references and number of citations per included article

## 3.2 Cross stages results

In this section we discuss the commonalities and differences across the four product development stages. For instance, Table 1A (Appendix) lists the ten studies that bridge at least two development stages. Three of them connect system design and implementation; for example, Krahe et al. (2020b) use point-cloud autoencoders to push early shape concepts straight into detailed CAD models, while Lin and Chiu (2017) transfer user sentiments mined from reviews into parametric design rules. On the other hand, four papers span design-to-integration trade-offs (Lambert et al. 2014; Akkaya et al. 2016; Bertoni et al. 2020; Booth and Ghosh 2023), and only two build a traceable thread from integration to validation (Ferreira et al. 2021; Lanus et al. 2021). Finally, only one paper mentioned both system implementation and integration stages (Mun et al.

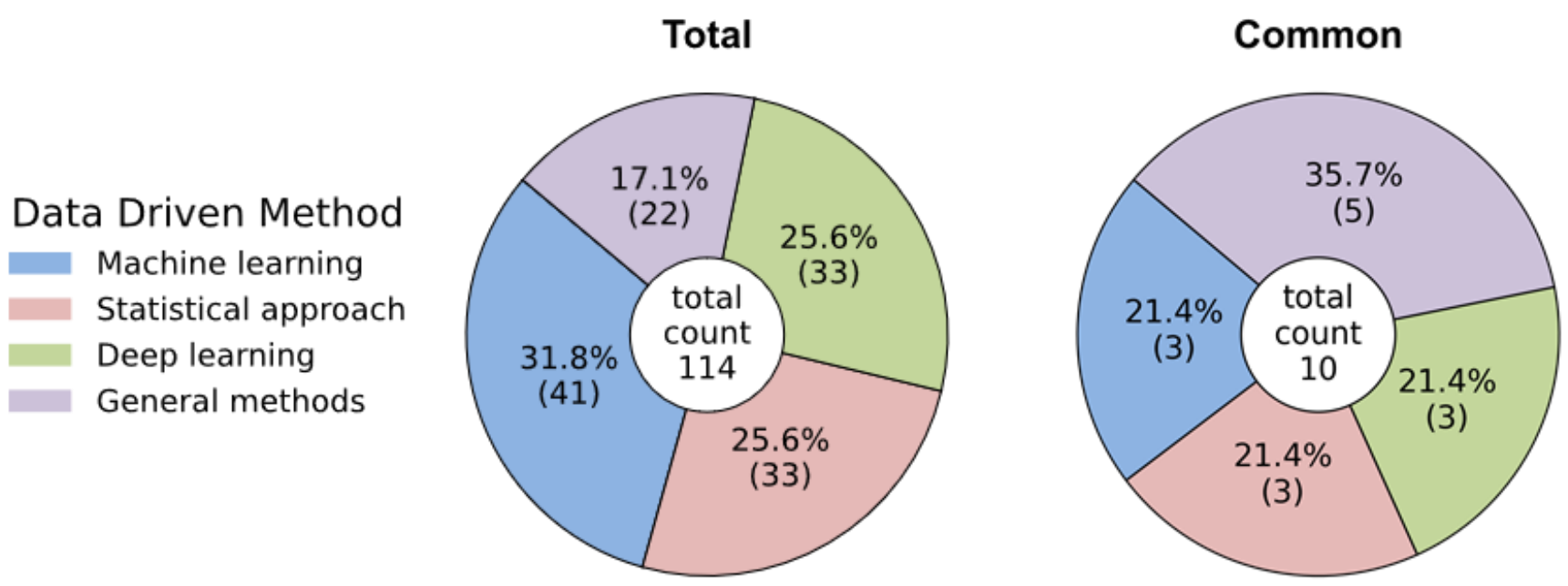

Figure 8: Results of the distribution of DDMs across product development stages in all included studies (left) and studies spanning multiple product development stages(right).

(2021). No study covers the full life-cycle of (Lambert et al. 2014) product development

Figure 8 illustrates the distribution of DDMs used in all included studies compared to those applied across multiple product development stages. The left chart shows that ML is the most dominant category, followed by statistical/probabilistic methods, DL, and General methods. In contrast, the right chart demonstrates a more balanced distribution across all four categories in cross-stage studies. This indicates that multi-stage applications tend to incorporate a broader mix of methods, likely due to the increased complexity and varied requirements of spanning multiple development phases. Furthermore, Figure 9 then breaks the 114 papers down by DDMs family: machine-learning dominates design and implementation, deep-learning gains traction in integration, while validation still relies mainly on classical statistics. Taken together, the figures illustrate a clear left-to-right decline in both publication volume and methodological diversity.

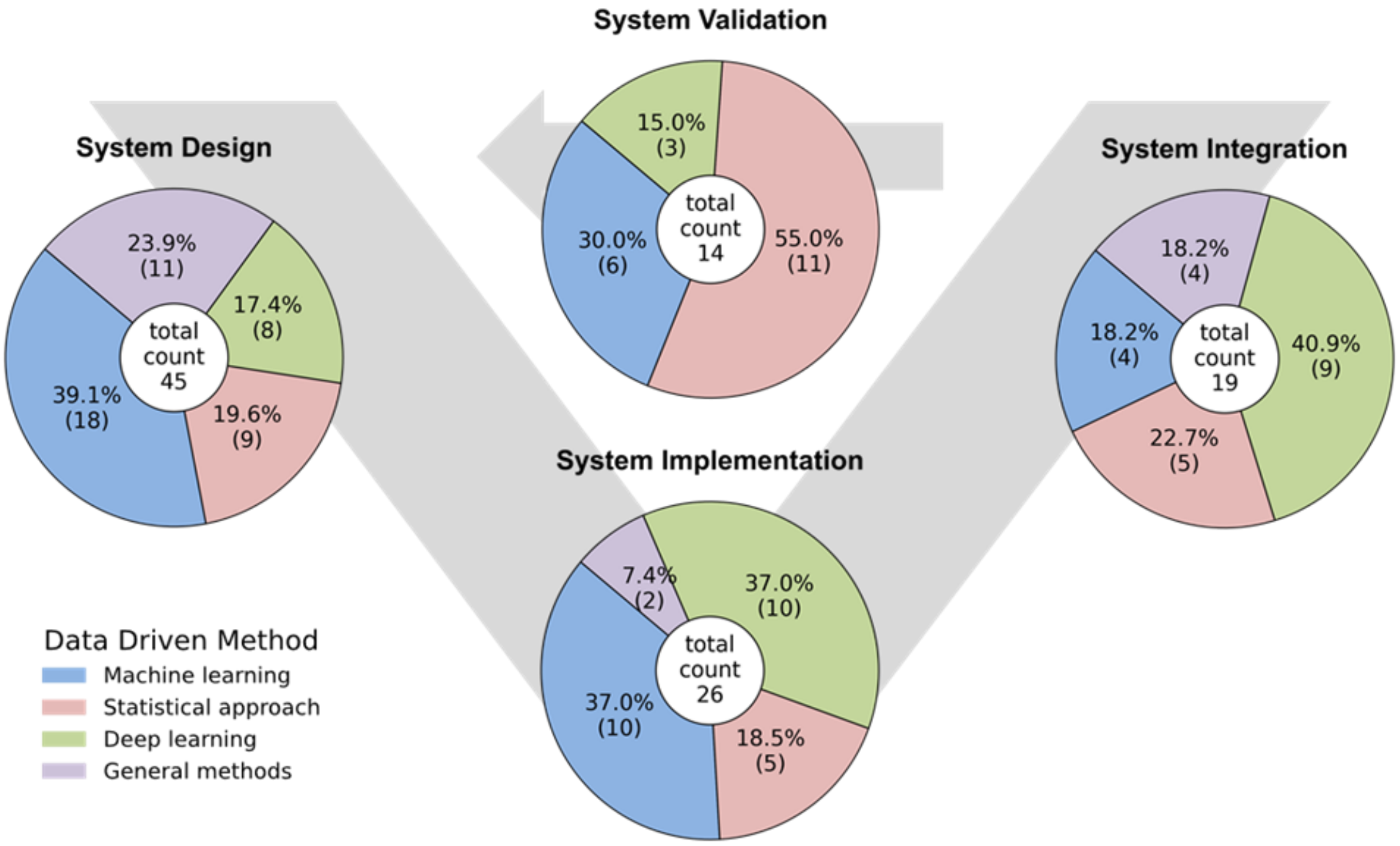

Figure 9: Results of the distribution of DDMs across the simplified V-model for product development stages namely; System Design, System Implementation, System Integration and Validation

Additionally, Figure 10 charts publication activity from 2014 to 2024 for each V-model stage. Output is modest and fairly even until 2020, after which a sharp rise occurs; peaking at eleven system-design papers in 2021 and mirrored by smaller increases in the other stages. System design stage consistently receives the highest number of publications throughout the observed period. System implementation maintains a moderate and steady presence, while System integration shows a notable rise, particularly after 2020, indicating emerging interest in applying DDMs to complex system-level interactions. Validation remains the least addressed stage, with slightly increased in recent years, highlight a potential underexplored area.

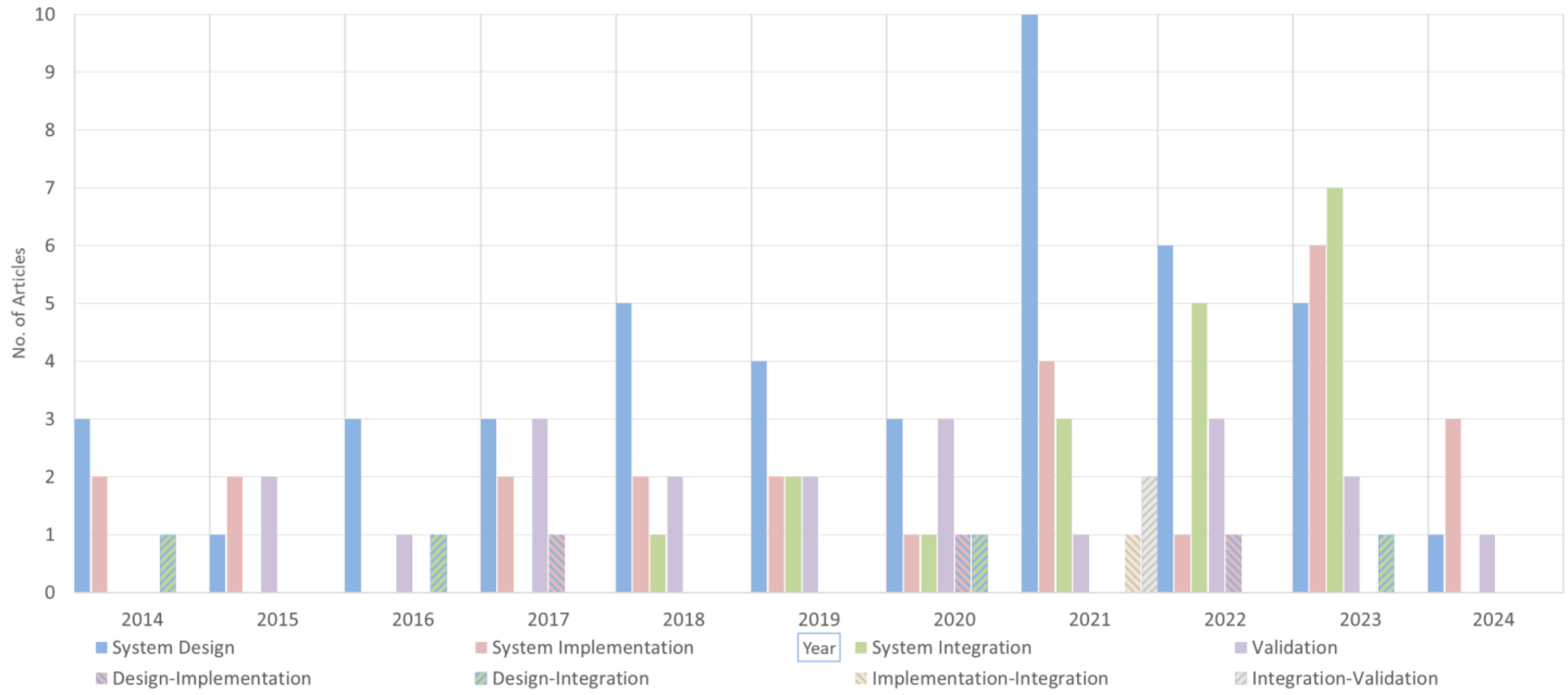

Figure 10: Distribution of included studies over time by product development stage and cross-stage classification (2014–2024).

To complement the method-stage mapping, Figure 11 illustrates the types of data used across product development stages. Numerical data is by far the most commonly used across all stages, especially in system implementation and validation. System design, integration and cross-stage studies show slightly more diversity, with some use of image and textual data. However, data types such as text and images remain underrepresented overall, and a small number of studies do not report the data type used. This suggests that DDMs in engineering design still rely heavily on structured numerical inputs, with limited exploration of multimodal or unstructured data sources.

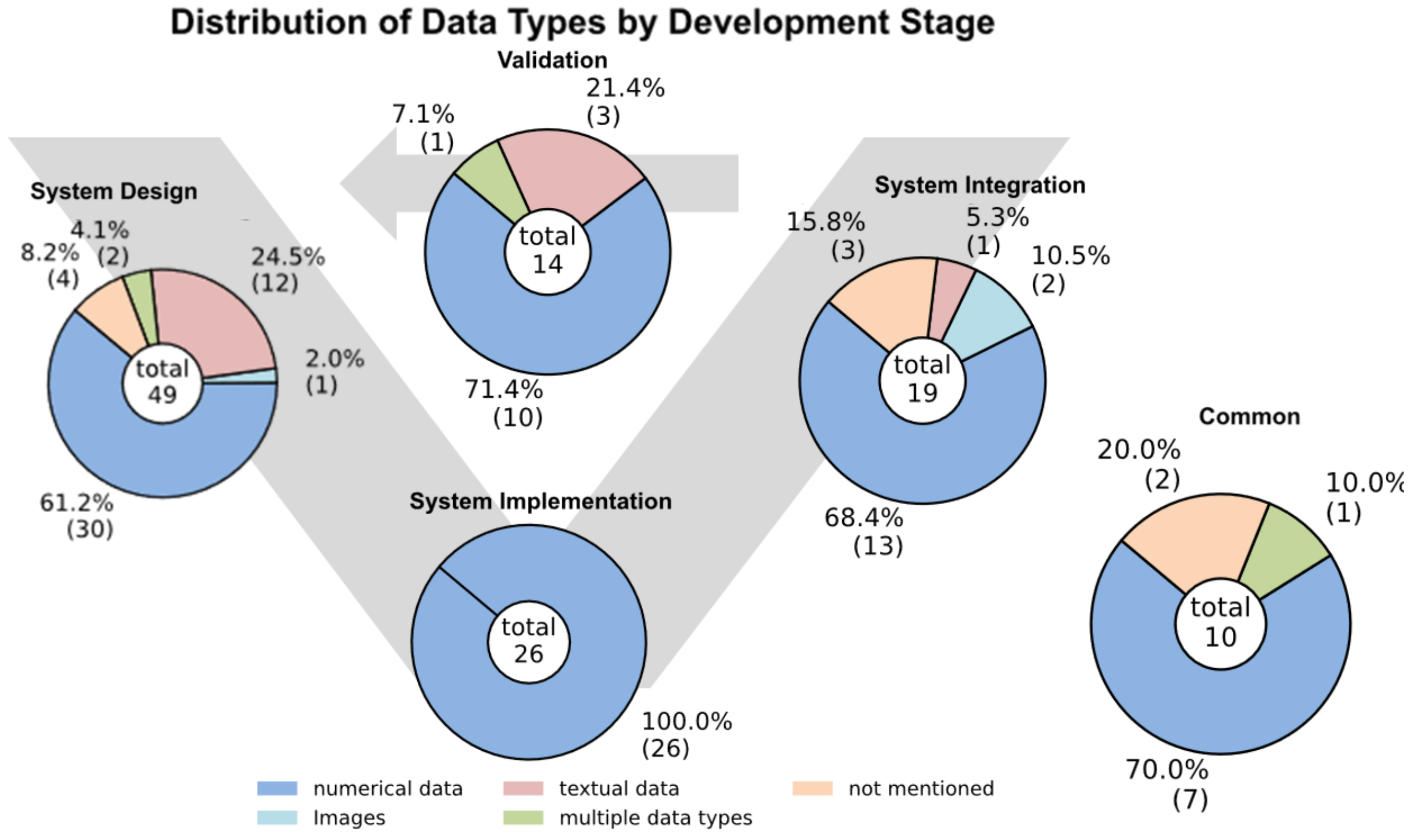

Figure 11: Results of DDM classification with respect to the different used datatypes

### 3.3 System Design stage

Based on the search string and the first screening, an initial set of 168 papers was identified. After the second screening, 116 papers were excluded as they didn't fit the screening criteria. This results in a final set of 52 papers, including 28 journal articles (54 %), 23 conference contributions (44 %), and one book chapter (2 %) see Figure 12. These publications are listed in Table A1 (Appendix), if they were also found in other phases, and in Table A2 (Appendix) if they were only found in the system design. Figure 12 also shows the distribution of the 52 included papers of system design between 2014 to 2024 their distribution. In which, it shows a steady number of contributions then peaked at 2021 with 11 articles.

The part of the V-model covered by system design contains activities that contribute to the decomposition of the system into implementable units. These activities include the comparison of alternative solution approaches with regard to various criteria, the inclusion of users in the consideration of solution approaches and the decomposition of the system from a functional, property and implementation perspective. This is reflected in the publications found. Looking at the applications in terms of recurring groups, the following emerge: 19 of the publications deal with the *evaluation of concepts*, for example Fusaro and Viola (2018); 5 publications deal with the *clustering of concepts*, for example Zhang et al. (2017); 5 publications deal with the *involvement of users*, for example Chiu and Lin (2018); 4 publications deal with the *automation of models*, for example Akay et al. (2021); and 4 publications deal with *design space exploration*, for example Bertoni et al. (2020). The remainder of the publications could not be grouped. The grouping of the publications shows that the comparison of different solution approaches in the form of concept in particular is addressed as an activity in accordance with the VDI 2206. While the inclusion of users is also apparent, decomposition is more of a subordinate activity, included for example through supporting the development of models, which in turn decompose the system. The evaluation of concepts is by far the most frequently identified application.

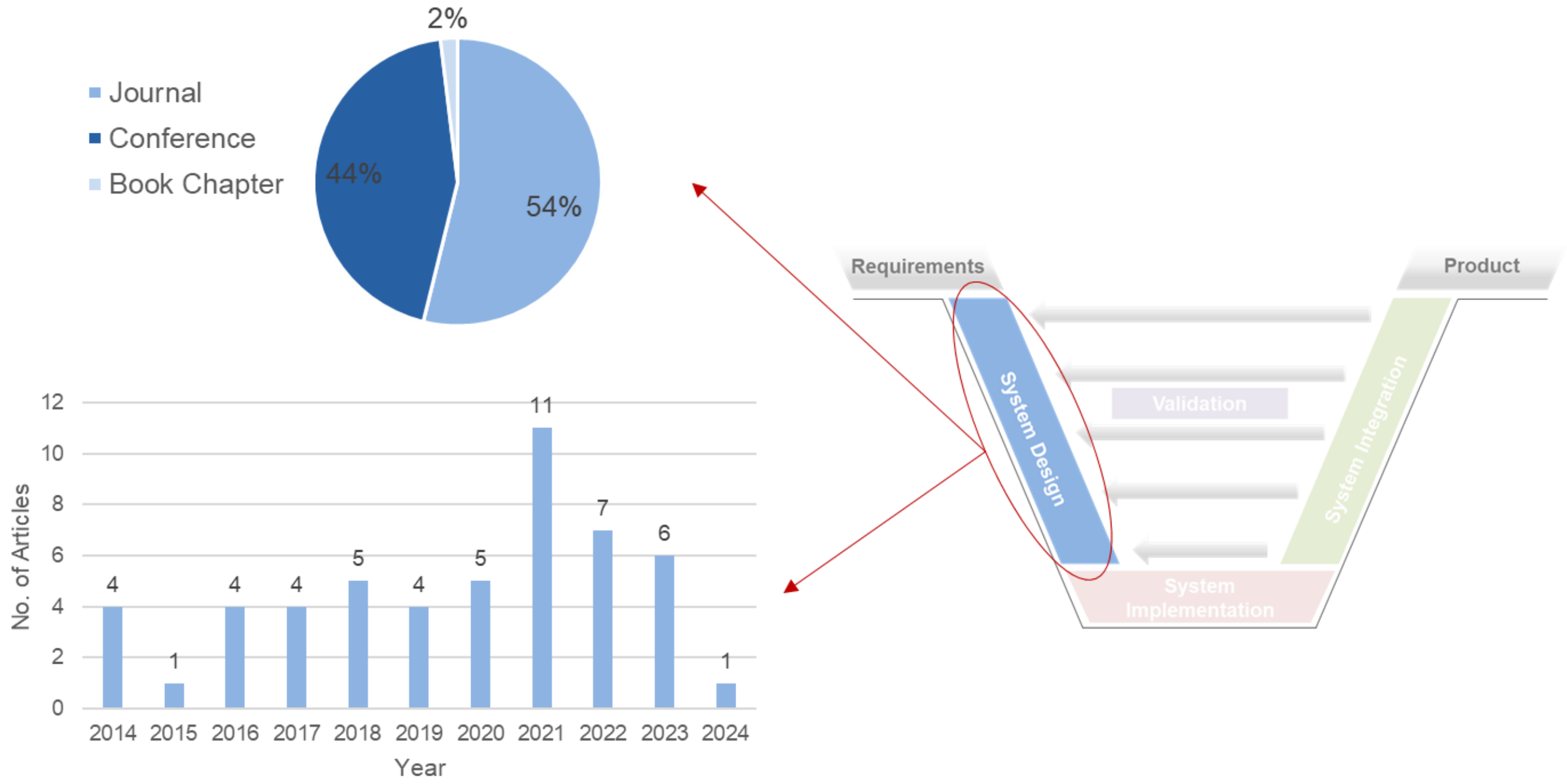

Figure 12: Distribution of publication types among included system design articles

In system design, 18 publications that use machine learning, 9 publications that use statistical approaches, 8 publications that use DL, 9 publications with general DDM approaches, and 2 reviews were identified. The general DDM approaches could not be assigned to any of the other approaches. This is due to the fact that these repeatedly refer to the use of a DDM in general, without explaining the approach in detail or assigning it to a higher-level approach. A look at the algorithms reveals hardly any multiple mentions. Only artificial neural networks and linear regression are more frequent with 4 cases each. Also, the algorithms identified are mostly types/classes of algorithms rather than explicit algorithms. The naming of a specific algorithm is a rarity.

In line with the ratio of approaches, by far the majority of types/classes of algorithms belong to ML. Mapping the data types to the DDMs shows that numerical data is processed 30 times, textual data 12 times, image data one time, and various data which could not be specified 2 times. The numerical data is in turn distributed 13 times to ML, 9 times to statistical approaches, 4 times to DL, and 5 times to general DDM approaches. The textual data is distributed 5 times to ML, 5 times to DL, and 2 times to a general DDM approach. The image data belongs to a general DDM approach. The various data belongs to 2 general DDM approaches and a combined use of machine learning and a statistical approach. In summary, ML approaches, which most often process numerical data, are preferred in the system design phase.

Looking at the limitations mentioned in relation to the approaches in the publications, there is a wide range: In the case of statistical approaches, computational effort (see Lambert et al. 2014) and specific boundary conditions of the application (see Umaras et al. 2021) are mentioned. For ML, the academic context of the development process (see Li et al. 2019a), large models and computing performance (see e.g., Papakonstantinou et al. 2014; Kreis et al. 2020), dependability on the input step files (see Dworschak et al. 2019), and the use of synthetic data (see Bertoni et al. 2020) are mentioned. For DL the unclear interaction with designers (see Akay et al. 2021), low accuracy (see Krahe et al. 2020b), and necessary manual steps (see Krahe et al.) are mentioned. Thereby, the limitations particularly relate to computational effort and challenges in relation to the quality of the input data. In terms of the type of studies conducted in the publications, in most cases, it is a case study (43 times), in which a mostly new approach is presented and then applied to a case. There are also a few field studies and studies with test subjects (5 times), and one more theoretical thought experiment. The focus in this phase is therefore strongly on the provision of new approaches and their presentation in the form of case studies in which considerable limitations are identified.

## 3.4 System Implementation stage

An initial set of 72 publications were identified based on the search string and the first screening. Following a second screening phase, 42 publications were excluded for not meeting the inclusion criteria. This resulted in a final selection of 30 relevant works, comprising 15 journal articles (50%) and 15 conference papers (50%). Furthermore, Figure 13 illustrates this as well as the the distribution of the 30 included between 2014 to 2024. Publications that appeared in multiple phases are listed in Table 1A (Appendix A), while those identified exclusively during the system implementation phase are listed in Table 4A (Appendix A). The following analysis is based on the system implementation publications included in both tables.

The analyzed papers were clustered according to the key activities of the system implementation stage: *geometric design, simulation, design optimization, and tolerancing*. A significant focus lies in *geometric design*, with 11 papers addressing various applications. Thukaram and Mohan (2019) utilized ML techniques to translate user input into product geometry. Aiming at accelerating the creation of CAD models, Krahe et al. (2020a) refined CAD automation by analyzing model trees to extract patterns in design feature sequences. A related but distinct application focuses on retrieving similar or fitting parts from existing CAD databases, Zhang et al. (2020) used a two-stage model (FilterNet + RankNet) to match user inputs; such as sketches and drawings; to suitable 3D geometries. In the *simulation* activity, three publications focused on replacing simulation methods with DDMs and accelerating simulation time and checking the models for plausibility. Askhatova et al. (2023) replaced conventional FEM-based topology optimization with convolutional neural networks (CNNs), drastically reducing computational time while maintaining comparable accuracy. Rastogi et al. (2023) applied statistical energy analysis and ML to model sound radiation, achieving a 1500× speedup compared to the boundary element method. *Design optimization* was the most frequently addressed activity, with 13 papers. 10 of these focused on optimizing specific performance parameters. Ding et al. (2022), e.g., forecasted fatigue failure in gear teeth using predictive models, allowing designers to adjust geometries accordingly. Li et al. (2015) used a Takagi-Sugeno fuzzy model to relate performance metrics to design parameters, enabling multidimensional design space exploration. Additional studies focused on accelerating the optimization process itself. Ciklamini and Cejnek (2024) integrated reinforcement learning into FEM simulations to shorten optimization cycles and adaptively explore design spaces. In the *tolerancing* activity the main use case identified is automated tolerance analysis and prediction, being addressed by two publications. Aschenbrenner and Wartzack (2017) statistically analyzed bearing seat tolerances in relation to operating clearance and fatigue life, helping to predict performance degradation.

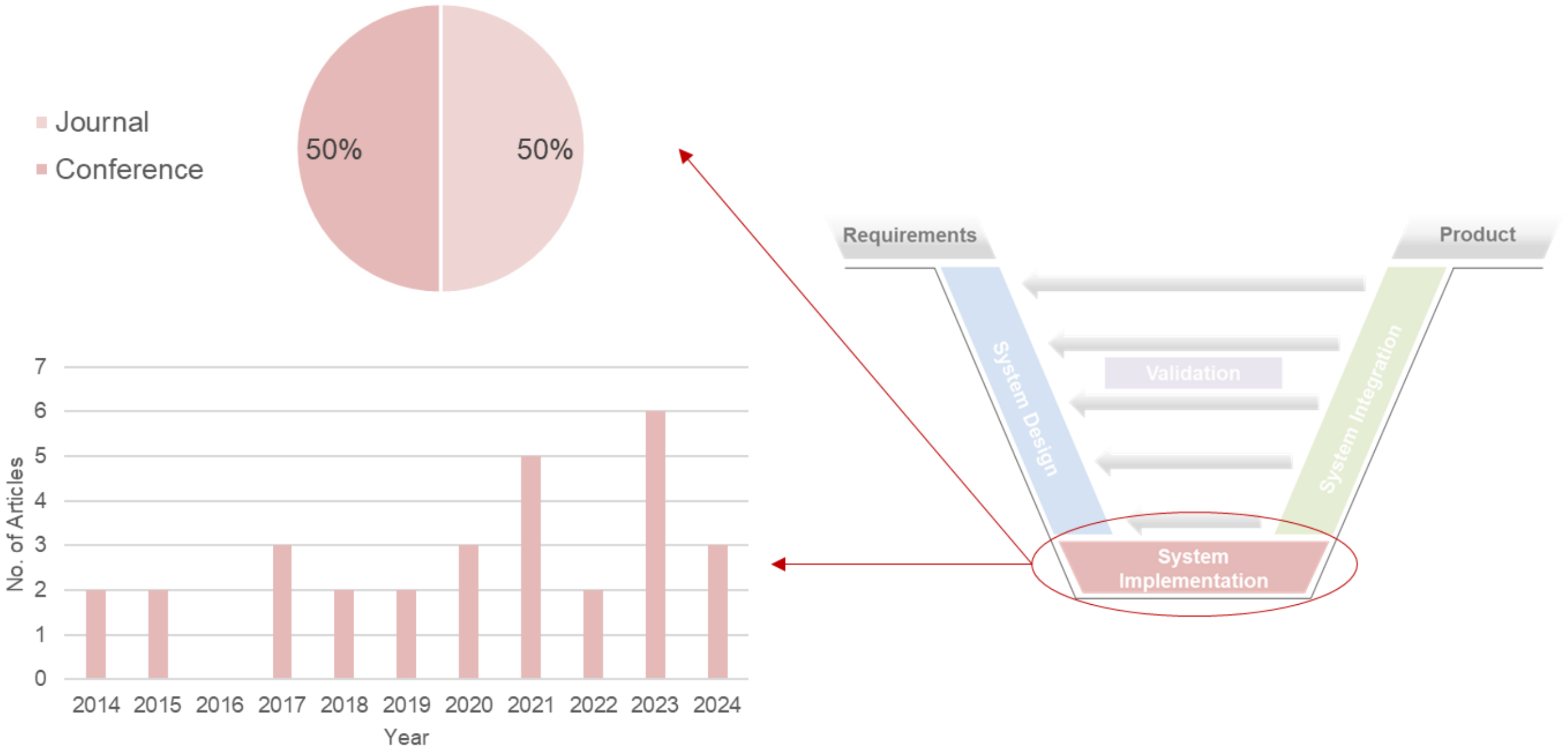

Figure 13: Distribution of publication types among included system design articles

In the implementation stage, ML (13 times) and DL (12 times) approaches are widely used due to data availability at this development stage, while statistical approaches (6 times) are less frequent. Some actually used hybrid DDMs (3 times) with more than one approach or compare results of ML and DL approaches. Among statistical approaches, Monte Carlo simulation (3 times) was most common, although other methods like Spearman Correlation or Statistical energy analysis were also utilized (1 time each). Prominent ML Algorithms include Support vector machine (SVM, 4 times), random forest (RF, 3 times) and K-Nearest Neighbors (kNN, 2 times); however, many alternatives were also explored, and results compared to the prominent ones, indicating uncertainty in optimal algorithm selection. DL predominantly featured neural networks (8 times). Generative Adversarial Networks (GAN, 1 time) were used specifically for geometry generation tasks.

Although mostly only numerical data (27 times) is used during the system implementation stage, a deeper analysis of the character of the numerical data is worthwhile, as it reflects the diverse requirements of DDMs in mechanical engineering design. Geometrical data, such as CAD models or other forms of geometric representation, appeared most frequently in the reviewed publications (20 times). Simulation data was used in 9 cases, supporting performance prediction and design optimization while reducing the need for physical testing. Numerical data from experiments was identified in 4 publications and is primarily used to validate models and ensure accuracy based on real-world measurements. Textual data, including customer reviews or Bills of Materials (BOM), was mentioned in 2 cases, one of them combined with numerical data. Images were used one time in combination with numerical data from CAD models.

Analyzing the limitation of the approaches it can be said that a common issue is the *limited scope of analysis* (10 times). Many contributions concentrate on isolated aspects of engineering problems, such as specific boundary conditions, thermal stress, or vibrational behavior, without addressing the full spectrum of influencing factors. Additionally, research is frequently confined to single-part models, with limited attention given to complex assemblies or inter-component dependencies. *Dimensional constraints* (5 times) also pose a challenge. A significant portion of the studies rely on two-dimensional (2D) representations to simplify computation and data handling. However, this restricts the transferability of findings to real-world three-dimensional (3D) applications, where geometrical complexity and spatial interactions play a critical role. Further, *computational and modeling simplifications* (5 times) are often introduced to reduce resource requirements. Surrogate models and reduced-order simulations are commonly used in place of full-scale numerical simulations, but this comes at the cost of accuracy and fidelity. These simplified approaches may overlook important nonlinear effects or coupled behaviors. Another limitation lies in *validation and generalization* (5 times). Many studies are evaluated using small datasets or under controlled conditions, which limits confidence in their applicability to broader engineering problems. Only a few papers incorporate large-scale or real-world validation, and the lack of such empirical testing hinders the assessment of robustness and reliability. Overall, the main challenge lies in the *limited generalizability* of current data-driven methods. This is largely a result of the simplified models and constrained datasets employed in the studies. Throughout the analyzed papers data-based approaches are predominantly used (22 times), followed by simulation approaches (8 times), while text-based approaches have not been identified. Looking at the research approach, in most cases it is a case study (25 times), in which a mostly new approach is presented and then applied to a case. Some works conducted field studies (5 times), utilizing experimental data from the field. The focus in this area is therefore strongly on the provision of new approaches and their presentation in the form of case studies.

## 3.5 System Integration stage

Following the initial search and first-stage screening, 47 papers were retained. The second-stage screening excluded 21 papers that did not meet the inclusion criteria, yielding a final sample of 26 papers: 10 journal articles (38.5%), 15 conference papers (57.7%), and 1 book chapter (3.8%) shown in Figure 14. Among these 26 papers, seven address multiple development stages: four engage with system design, one with system implementation, and two with validation as shown in Table 1A (Appendix A). Adding to this, system integration publications were sparse prior to 2020, then rose sharply from 2 articles (2020) to 6 (2021) and peaked at 9 in 2023, indicating accelerating interest in the topic as shown in Figure 14. The single article in 2024 likely reflects an incomplete publication year or a short-term lull rather than a reversal upward trend.

Based on the literature analysis of 26 papers, various applications of DDMs in the context of system integration could be identified. The reviewed papers were categorized according to major integration activities defined by (VDI/VDE 2206, 2021): *verification, subsystem coordination and interaction modeling, system behavior prediction, test prioritization, and integration architecture design*. Verification were addressed in eight studies, where DDMs support test case prioritization, test sequencing, and simulation-based verification processes. (Ferreira et al. 2021) presented a hardware-in-the-loop (HIL) testing prioritization approach that applies statistical learning techniques to optimize test sequencing for embedded system integration. (Lanus et al. 2021) proposed a simulation framework for multi-agent systems, where scenario-based evaluation helps identify critical system interactions during verification. On the other hand, subsystem interaction modeling was addressed in studies that explored coordination and verification dependencies. (Wöhr et al. 2023) examined integration and verification complexity through control system modeling and coordination experiments. (Booth and Ghosh 2023) contributed a model documentation framework to better integrate ML models into system engineering artifacts, addressing traceability and consistency across subsystems.

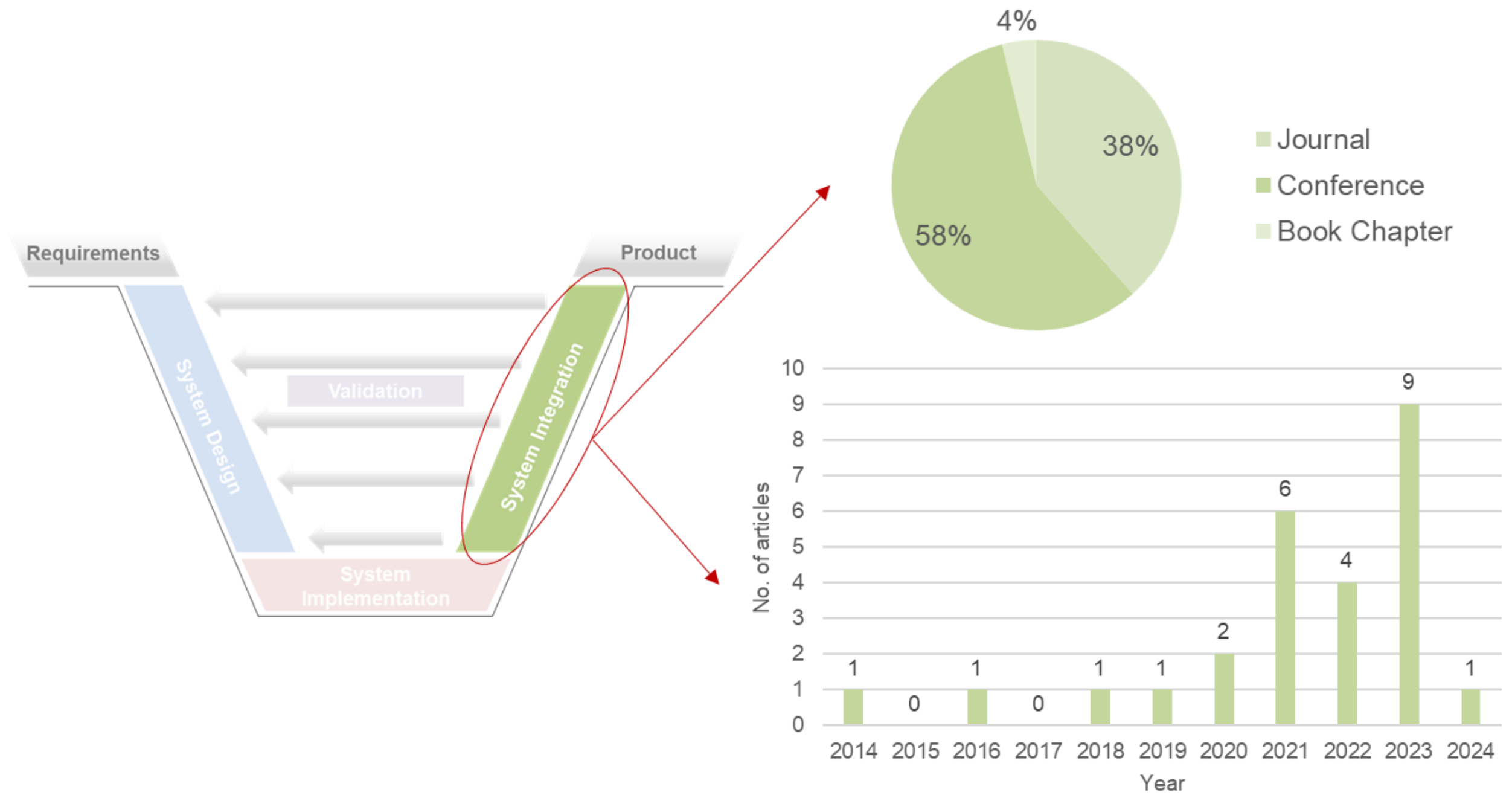

Figure 14: Distribution of publication types among included system integration articles

System behavior prediction was a frequent topic, where machine learning, DL, and hybrid methods were used to predict system-level behavior, failure likelihood, or remaining useful life (RUL). (Rosen and Pattipati 2023) employed physics-regularized models for health prognostics of complex engineered systems with dependent subsystems. Similarly, (Synnes and Welo 2023) demonstrated the use of physics-informed neural networks (PINN) to support prognostics and predictive maintenance in digital twin environments.

Integration coordination and architectural design were supported by data-driven tools to optimize early integration activities. (Gay et al. 2021)) presented knowledge integration approaches to include expert knowledge into early health management design. (Akkaya et al. 2016) applied aspect-oriented modeling for early system architecture design in cyber-physical systems, allowing consideration of cross-cutting concerns early during integration.

From a methodological perspective, ML remains the most applied DDM (11 publications), followed by DL (5 publications), statistical/probabilistic methods (5 publications), as well as physics-informed models and hybrid approaches (5 publications). General data-driven frameworks were applied where integration scenarios involved mixed simulation, knowledge models, and heuristic coordination (e.g. (Niu et al. 2018; Su et al. 2019)). In terms of algorithms, the studies demonstrated considerable variety. Support Vector Machines (SVM), ensemble models, Bayesian networks, deep autoencoders, multitask learning, particle filtering, and expert knowledge fusion methods were used for failure prediction, health prognostics, coordination analysis, and multi-agent system evaluation (e.g., (Lambert et al. 2014; Bertoni et al. 2020; Janson et al. 2022; Dizor et al. 2023; Zhang et al. 2024)). Simulation-based algorithms such as reliability-based design optimization (RBDO), Monte Carlo simulations, and scenario generation were widely adopted in uncertainty propagation and verification optimization (e.g., (Hajiha et al. 2022; Gamage et al. 2023)).

With respect to data types, numerical time series data from sensor monitoring and condition monitoring appeared in 13 publications, simulation data was used in 10 publications, and textual data, including system models and documentation, appeared in at least 3 studies (Li and Liu 2022; Booth and Ghosh 2023). In this integration stage, geometric data appeared less frequently compared to design and implementation stages, since the main challenges relate to system-level behavior rather than geometric representations.

The reported limitations of DDMs in system integration largely reflect several recurring technical barriers: limited availability of rare-event or failure data for training predictive models, lack of generalizability of learned models to new operating conditions, challenges in coordinating verification processes across multiple interacting subsystems, and high computational requirements for uncertainty quantification (Lambert et al. 2014). In addition, the incorporation of machine learning into formal system engineering processes remains challenging due to documentation gaps, model traceability, and integration consistency (Booth and Ghosh 2023). The majority of studies employed case study methodology (22 publications), where new DDMs were demonstrated on representative integration problems. A smaller portion (4 studies) included real-world field data or operational testing. This reflects the exploratory nature of system integration research, where practical challenges are addressed and tested through representative, but often limited, case studies.

### 3.6 Validation stage

Based on the search string and the first screening, an initial set of 42 papers was identified. After the second screening, 27 papers were excluded because they did not address concrete data-driven validation methods. This results in a final dataset of 15 papers, including 6 journal articles (40%) and 9 conference contributions (60%) shown in Figure 15. Additionally, shows the distribution of the 30 included between 2014 to 2024.

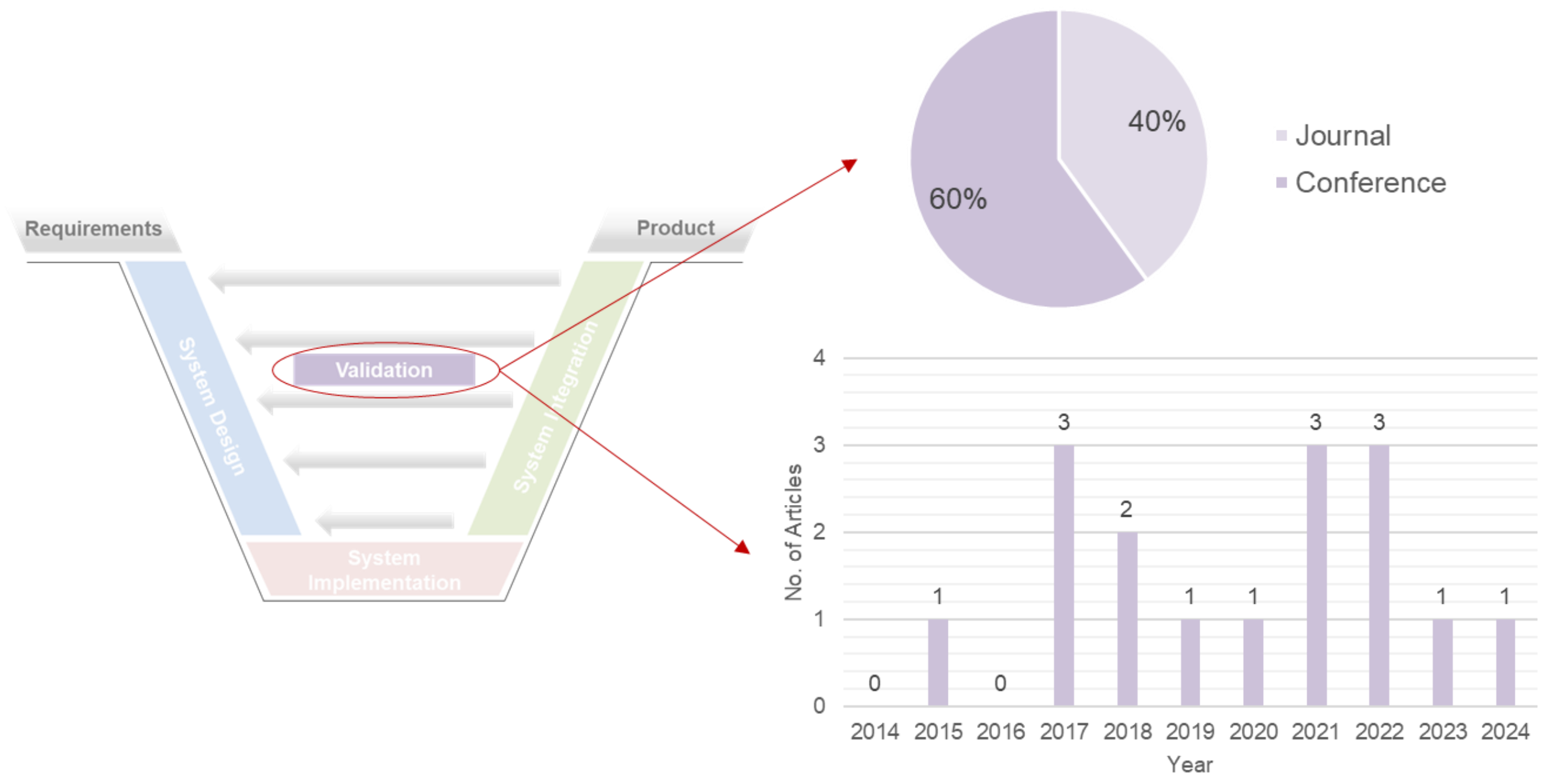

Figure 15: Distribution of publication types among included system design arttircles

Different types of data are used depending on the study focus and methodology. While quantitative data dominates due to its suitability for statistical analysis and modeling, qualitative data can provide valuable contextual insights. Our review shows the following distribution: 17 studies use quantitative data, and 4 study uses qualitative data. Validation can be performed at different integration levels. While the system level represents the highest integration level, subsystems and individual components can also be subject to specified validation. Our review reveals the following distribution: 6 studies at the system level, 12 at the subsystem level, and 4 at the component level. The domain of the investigated systems varies in terms of their implementation context. Validation can be carried out in virtual environments, on physical systems, or through a combination of both. Our review reveals the following distribution: 6 studies focus on virtual systems, 7 on physical systems, and 6 employs a mixed approach like Hardware-in-the-Loop. Statistical methods play a fundamental role in system validation, particularly where analytical modeling and uncertainty quantification are required. Kriging-based surrogate modeling is frequently used to approximate computationally expensive simulations by creating statistically sound interpolators with associated confidence bounds, enabling efficient exploration of design spaces (Son et al. 2020). Statistical degradation modeling supports lifetime estimation by capturing time-dependent performance loss through probabilistic trend functions, often used in reliability evaluations (Chen et al. 2024). Noncausal finite impulse response (FIR) modeling in the frequency domain allows for the identification and validation of dynamic system behavior, providing a spectral representation of input-output relationships critical for understanding complex signal responses (Zheng et al. 2023). Hypothesis tests (t-Test, Wilcoxon, Chi-Square) are fundamental statistical methods for evaluating differences or dependencies between groups or variables, supporting data-driven validation processes. Weibull, Normal, and Lognormal distribution fitting is a standard approach to model lifetime data and failure probabilities in reliability evaluations (Jung et al. 2015). Maximum Likelihood Estimation is used to optimize parameters for parametric models, allowing for accurate statistical modeling in validation tasks (Sekar et al. 2022). Bayesian networks provide probabilistic modeling capabilities by representing dependencies between variables and quantifying uncertainties in system validation, as demonstrated by (Douthwaite and Kelly 2017; Rizzo and Blackburn 2019) and (Rashid et al. 2015). Decision trees and k-nearest neighbor classifiers offer interpretable rule-based structures for classification tasks in system validation, enabling robust scenario classification and specification alignment. Deep Residual Recurrent Neural Networks (DR-RNN) integrate deep residual networks with recurrent structures to model dynamic and temporal dependencies in system validation, as implemented by Liu and Goebel (2018). Graph Neural Networks (GAT) capture complex relational structures in graph-based system data, aiding in validation of networked system behaviors, as demonstrated by Kyamakya et al. (2022b). Signal Temporal Logic is a formal verification approach to validate time-dependent logical requirements for system behavior, supporting rigorous system-level validation processes. Exploratory Data Analysis (EDA) is used to reveal reliability trends and identify potential anomalies in validation data, providing important context for data-driven decision-making (Kyamakya et al. 2022b).

Machine learning algorithms are widely adopted in validation workflows to uncover patterns in high-dimensional datasets and support data-driven decision-making. Classification-tree approaches and decision tree algorithms offer interpretable rule-based structures for scenario classification and specification alignment, enabling traceable validation processes (Bach et al. 2017b). Clustering algorithms, such as k-means or hierarchical clustering, are utilized to group similar operational states, improving representativeness and reducing test redundancy. ML regression models with k-fold cross-validation enhance predictive validation by learning continuous relationships from empirical data while providing internal performance estimates (Petersen et al. 2019). Information Theoretic Metric Learning (ITML) improves validation in distance-based applications by learning task-specific similarity metrics that better reflect system behavior differences (Sohier et al. 2021). Beyond purely statistical or ML methods, hybrid and modeling-oriented tools provide essential capabilities for system-level validation. The Markov Decision Process (MDP) is employed as a framework for modeling sequential decision-making under uncertainty, particularly relevant in reinforcement learning-based validation strategies where systems must adapt to dynamic environments (Ellis et al. 2022).

Validation predominantly relies on numerical data, which includes both continuous and discrete measurements such as temperature, pressure, vibration, or time-series sensor signals. This type of data enables precise quantification of system behavior and is essential for statistical analysis, modeling, and algorithmic validation tasks. Its structured nature allows for consistent preprocessing, feature extraction, and integration into ML and simulation-based approaches. In contrast, non-numerical data types such as textual data (e.g., formal specification, …), or image data (e.g., visual inspections) are rarely used in the analyzed literature. Although such data could enrich system understanding and support complementary validation tasks, such as anomaly detection through images, they remain underrepresented.

The analysis of the 15 selected publications reveals four primary application areas for data-driven validation approaches: Reliability validation, performance validation, safety validation, and operational validation. Reliability validation (2 papers) focuses on predicting and preventing failures. Representative approaches include text mining-based methods to support Design Failure Mode and Effects Analysis (DFMEA), such as automated extraction of failure causes and effects from historical defect descriptions and classification of failure relationships using natural language processing (Li and Wu 2018), as well as remaining useful life estimation for oil pumps (Chen et al. 2024). Performance validation (5 papers) aims to ensure that systems meet their functional requirements. Examples include the prioritization of hardware-in-the-loop (HiL) tests using data analytics (Ferreira et al. 2021), statistical model validation based on limited experimental data (Jung et al. 2015), comparison of driving simulator outputs with real-world vehicle data (Sekar et al. 2022), energy consumption and range estimation for electric vehicles with consideration of driver-specific and driver-unspecific factors (Petersen et al. 2019), and vibrational behavior estimation for automotive steering columns (Son et al. 2020). Safety validation (3 papers) addresses challenges in safety-critical systems. These include the establishment of verification and validation (V&V) criteria for Bayesian networks in safety-relevant applications (Douthwaite and Kelly 2017), the use of predictive cruise control systems for geolocation-specific test scenarios (Bach et al. 2017b), and safety validation strategies for autonomous systems (Ellis et al. 2022). Operational validation (4 papers) focuses on assessing system performance under real-world conditions. Approaches include the application of deep learning for prognostics in the national airspace system (Liu and Goebel 2018), continuous data collection in automotive engineering environments (Bach et al. 2017b), and signal-based estimation of vehicle body accelerations (Zheng et al. 2023), and real-time anomaly detection in smart patient monitoring systems, combining sensor fusion, rule-based filtering, and adaptive modeling to capture deviations in weight, movement, and bed presence (Kyamakya et al. 2022a).

There are five main categories of limitations of DDMs in validation emerged from the analysis: Data availability and quality, scalability, model interpretability, domain specificity, and computational constraints. Data availability and quality (9 papers) are often constrained by the rarity of failures and incomplete field data, as noted by (Kyamakya et al. 2022a). Approaches are also limited by low data diversity and test coverage (Petersen et al. 2019). Additionally there can be a strong dependency on training data (Zheng et al. 2023). Concerning scalability (5 papers), the growing system complexity leads to an exponential increase in required tests (Lanus et al. 2021). Due to the black-box nature of many ML approaches, their lack of interpretability, as discussed in five papers, poses a challenge for deployment in safety-critical applications. It is also constrained by simplifications (Son et al. 2020; Sohier et al. 2021). As highlighted in six papers, many methods are tightly coupled to specific domains, which limits their broader applicability and generalizability, e.g. (Son et al. 2020). Across six papers, computational constraints, particularly in embedded systems, highlight the challenge of balancing thorough validation with the need for real-time performance. Limitations are also caused by semi-automation of the processes (Sohier et al. 2021), high costs (Son et al. 2020) and resource requirements (Zheng et al. 2023).

# 4. Discussion

This section discusses the findings of the systematic literature review in relation to the central research question: ***What is the current state of application, challenges, and opportunities in applying data-driven methods (DDMs) across the product development lifecycle of mechanical and mechatronic systems?*** It answers the three sub-questions (SQ) by reflecting critically on the results, placing them within the context of existing literature, and then presenting implications for future research.

## 4.1 SQ 1: How are DDMs implemented in engineering product development, and what data types and algorithms do they rely on?

The review confirms that ML and statistical approaches are the most prevalent DDMs employed in mechanical and mechatronic product development. These findings confirm prior literature reviews (e.g., Ma & Wang, 2022; Fann et al., 2021) that highlight the dominance of supervised learning and regression-based statistical modeling in engineering applications. However, this study extends the current state of research by confirming and offering a stage-specific mapping of DDMs across the product development lifecycle, which previous reviews have only partially addressed.

Furthermore, supervised learning techniques such as decision trees, support vector machine (SVM), and ANN are commonly used in system implementation and integration stages, especially when imaging or simulation data are available. In contrast, unsupervised learning, including clustering and dimensionality reduction, appears most frequently in the system design stage, where it supports tasks like early-stage design space exploration, requirement structuring, and user segmentation.

DL methods, on the other hand, are present but less widespread. Their usage is primarily observed in geometry-intensive and sensor-based applications, such as image-based diagnostics or condition monitoring. The limited adoption of DL is likely due to the high data requirements and reduced interpretability of such models. These two barriers of DL have also been highlighted in other studies like (Zhang et al., 2020; Wuest et al., 2016), and the review results confirm that there is a lack of available data in system design stage which is challenging to implement and benefit from the potentials of DL methods.

Statistical approaches continue to be widely used, particularly in the implementation and validation stages. Linear and nonlinear regression, Monte Carlo simulation, and response surface modeling are applied for tolerance optimization, reliability analysis, and uncertainty quantification. Compared to ML models, statistical approaches are shown to be often favored in the validation stage due to their transparency and traceability, critical features for risk-sensitive engineering decisions. This supports earlier claims from reliability-focused design literature (Lyu et al., 2020) and underscores a persistent methodological divide between early exploratory phases and final evaluation stages.

Following on this, hybrid methods are also strongly emerging, combining statistical with ML methods. For example, surrogate modeling techniques often rely on regression-based metamodels (e.g., polynomial response surfaces) integrated with optimization algorithms to reduce the computational burden of high-fidelity simulations. These techniques appear most often in system design and integration, where they offer a compromise between model accuracy and resource efficiency.

## 4.2 SQ 2: How are these DDMs applied across the stages of the product development process?

DDM adoption is most mature in the system implementation stage, where a wide range of engineering activities, optimization, simulation support, and geometry generation, benefit from available simulation and CAD data. Supervised ML and statistical models are well-established in this stage, supporting tasks such as surrogate modeling and performance prediction.

System design demonstrates increasing use of DDMs, particularly for early decision-making. However, applications in this stage are generally exploratory, and limited by the availability of usable design data. The prevalence of unsupervised learning here reflects the less structured nature of early design tasks. DL adoption in design is minimal due to the scarcity of high-quality, labeled datasets.

Integration and validation stages are less represented in the reviewed literature. Integration-focused studies typically involve simulation-driven test planning or system behavior modeling, often limited to a single domain. Validation is characterized by conservative method use, with statistical models dominating due to their interpretability and ease of integration into risk-sensitive environments.

Overall, DDM use tends to decline in later stages of the V-model. The transition from design to system-level validation is marked by reduced algorithmic diversity and fewer studies demonstrating full lifecycle continuity. This fragmented application limits the broader potential of DDMs in supporting traceability and closed-loop design workflows.

## 4.3 SQ 3: What challenges limit, and what opportunities enable, the effective use of DDMs across the product development lifecycle?

### 4.3.1 Challenges

Challenges are both technical and methodological and vary across development stages. In early-stage design, the lack of structured data and the difficulty of linking ML model outputs to design decisions reduce the practical utility of DDMs. Many studies rely on illustrative or conceptual cases, limiting real-world generalizability.

In implementation, the key challenges are poor model reusability and the absence of benchmarking standards. Although geometric and simulation data are more available, datasets are often inconsistent or proprietary, hindering replication. Additionally, strong dependence on traditional physics-based simulations (e.g., FEA) continues to slow broader ML adoption.

Integration tasks suffer from data heterogeneity and a lack of system-level validation. Studies often remain limited to simulation environments without deployment in real operating contexts. Validation presents the most conservative stage, where model opacity and low data availability make ML adoption difficult. Statistical models dominate due to their traceability and low computational cost.

These findings support and extend the observations by Gerschütz et al. (2023b), who highlighted the fragmented and uneven maturity of DDM applications across the V-model.

### 4.3.2 Opportunities

The review identifies several areas where DDMs hold strong potential. First, hybrid approaches that combine data-driven and physics-based models are gaining interest, particularly in simulation-heavy tasks where empirical data alone may not suffice. These approaches can offer higher model fidelity while maintaining interpretability.

Second, design automation through generative models, surrogate models, and rule-based ML systems presents a clear opportunity to reduce time-to-design and support rapid iteration. While most work remains focused on conceptual design, there is scope to extend these methods to detailed CAD, tolerance analysis, and manufacturability evaluation, enhancing design-to-implementation continuity.

Third, implementation-stage applications show strong momentum. The availability of simulation, geometric, and image-based data enables a wide range of use cases, from part classification to surrogate-assisted optimization. Many of these methods aim to offload repetitive tasks, freeing engineers to focus on higher-value decision-making.

Finally, emerging applications in system integration and validation, such as explainable ML for fault detection, test prioritization, and health monitoring, indicate growing interest in expanding DDM use into system-level operations. Models that are interpretable, computationally efficient, and easily integrated into CAE/CAD environments demonstrate greater industrial applicability.

## 4.4 Implications, limitations, and future work

This review responds directly to the identified research gap by delivering the first stage-specific synthesis that links DDM types, algorithms, data, application tasks, and limitations across mechanical and mechatronic product development. Unlike prior reviews (e.g., Figoli et al. 2025; Payrebrune et al. 2024), which offered bibliometric or high-level surveys, this study offers detailed classifications that can serve as the first building block for future guidance.

However, this is only a partial solution. The current study does not offer prescriptive recommendations on which method to use when. Such guidance would require a second synthesis combining this review’s engineering-stage-specific findings with comprehensive taxonomies of data-driven algorithms from computer science and AI literature.

Therefore, future research should aim to combine engineering-stage mappings with algorithm-centric overviews from data science to build a dual-perspective guideline. Develop stage-aware decision frameworks that align DDM types with engineering tasks, data characteristics, and desired outputs. Establish community standards for reporting, benchmarking, and reproducibility in DDM applications. Ultimately, unlocking the full potential of DDMs in engineering design will require both technical advances and methodological structuring, enabling systematic deployment across the entire product development lifecycle.

By answering the 3 sub-questions. A final conclusion regarding the overarching question: What are the challenges and opportunities in applying DDMs to the product development of mechanical and mechatronic systems? is discussed. In summary, ML and statistical methods are the primary tools driving the data-informed transformation of product development in mechanical and mechatronic systems. Their applications are stage-specific: statistical methods dominate tolerance design, uncertainty analysis, and simulation support, while ML methods are leveraged for classification, pattern recognition, and optimization. Despite this breadth, current applications remain largely task-specific and lack full lifecycle continuity. The main limitations are: incomplete validation under real-world constraints, low methodological transparency and reproducibility, and weak integration with physical models and engineering workflows. Emerging trends point toward more hybrid models, interpretable learning, and broader integration via digital twin architectures. Realizing the full potential of DDMs requires not only technical improvements but also structured frameworks for their deployment across the entire V-model, ensuring continuity from concept to validation.

Analyzing the literature highlights a future need for DDMs guidelines to support engineering designers in adopting and benefiting from these methods. In which some barriers in having such guideline are discussed. For instance, the adoption challenge which includes: lack of accessible, high-quality data throughout the product development lifecycle, absence of standardized evaluation metrics and benchmarks, model opacity, particularly for complex ML architectures, and fragmented toolchains that inhibit integration into existing workflows. On the other hand, studies that explicitly address interpretability, computational cost, and integration with established engineering tools (e.g., CAD/CAE platforms) show higher transferability potential. Conversely, methods requiring extensive preprocessing or customization are less likely to be adopted in industrial settings.

## 5. Conclusion

This study presented a systematic literature review of DDMs applied in the product-development of mechanical and mechatronic systems. By mapping 114 studies to the V-model’s four simplified product development stages, system design, system implementation, system integration, and validation, the review provides a structured overview of the types of DDMs currently in use, the data and algorithms they rely on, the stage-specific challenges they face, and the opportunities they present. The findings show that machine learning and statistical methods dominate current applications, but their adoption is uneven across development stages. While system implementation exhibits the highest maturity, driven by simulation and geometric data, system design remains fragmented, and system integration and validation are underexplored. The review also highlights persistent challenges related to data quality, model transparency, and reproducibility, while identifying promising directions in hybrid modeling, design automation, and interpretable AI. In addressing the research gap, this review offers a foundational step toward the development of practical guidelines that support engineering designers and researchers in selecting and deploying DDMs effectively. However, it does not provide a prescriptive framework. To move toward that goal, future work should integrate these findings with complementary reviews from the data science domain that categorize algorithms by their learning strategies, input requirements, and application constraints. Such synthesis would enable the creation of comprehensive, stage-aware guidelines tailored to engineering activities and data availability, ultimately contributing to the more systematic and effective use of DDMs across the entire product development lifecycle.

## References

Ahmed F, Chen W (2023) Investigation of steam ejector parameters under three optimization algorithm using ANN. Applied Thermal Engineering 225. https://doi.org/10.1016/j.applthermaleng.2023.120205

Akay H, Yang M, Kim S-G (eds) (2021) Automating design requirement extraction from text with deep learning, Proceedings of the ASME Design Engineering Technical Conference, 3B-2021. https://doi.org/10.1115/DETC2021-66898

Akkaya I, Derler P, Emoto S, Lee EA (2016) Systems Engineering for Industrial Cyber–Physical Systems Using Aspects. Proceedings of the IEEE 104:997–1012. https://doi.org/10.1109/JPROC.2015.2512265

Allagui A, Belhadj I, Plateaux R, Hammadi M, Penas O, Aifaoui N, Choley J-Y, Allagui A, Belhadj I, Plateaux R, Hammadi M, Penas O, Aifaoui N, Choley J-Y (2023) Towards a Data-Driven Smart Assembly Design: State-of-the-Art. In: Springer Verlag (ed) Lecture Notes in Mechanical Engineering, pp 343–352. https://doi.org/10.1007/978-3-031-23615-0_35

Aphirakmethawong J, Yang E, Mehnen J (2022) An Overview of Artificial Intelligence in Product Design for Smart Manufacturing. In: 2022 27th International Conference on Automation and Computing (ICAC), pp 1–6. https://doi.org/10.1109/ICAC55051.2022.9911089

Aschenbrenner A, Wartzack S (2017) A method for the tolerance analysis of bearing seats for cylindrical roller bearings in respect to operating clearance and fatigue life. In: Proceedings of the 21st International Conference on Engineering Design (ICED17): Design methods and tools. Curran Associates Inc, Red Hook, NY

Askhatova A, Y. Gabdulla, A. Bekbolat, E. Shehab, M. H. Ali (2023) Machine Learning-Based Topology Optimization in 3D Printing. In: 2023 International Conference on Smart Applications, Communications and Networking (SmartNets), pp 1–6. https://doi.org/10.1109/SmartNets58706.2023.10216161

Ayensa-Jiménez J, Doweidar MH, Sanz-Herrera JA, Doblaré M (2019) An unsupervised data completion method for physically-based data-driven models. Comput. Methods Appl. Mech. Eng. 344:120–143. https://doi.org/10.1016/j.cma.2018.09.035

Bach J, Langner J, Otten S, Holzäpfel M, Sax E (2017a) Data-driven development, a complementing approach for automotive systems engineering. In: 2017 IEEE International Systems Engineering Symposium (ISSE), pp 1–6. https://doi.org/10.1109/SysEng.2017.8088295

Bach J, Langner J, Otten S, Sax E, Holzapfel M (2017b) Test scenario selection for system-level verification and validation of geolocation-dependent automotive control systems. In: IEEE (ed) International Conference on Engineering, Technology and Innovation (ICE/ITMC), pp 203–210. https://doi.org/10.1109/ICE.2017.8279890

Bachmayer R, Kampmann P, Pleteit H, Busse M, Kirchner F (2020) Intelligent Propulsion. In: Intelligent Systems, Control and Automation: Science and Engineering, vol 96, pp 71–82. https://doi.org/10.1007/978-3-030-30683-0_6

Batliner M, Boës S, Heck J, Meboldt M (2022) Linking Testing Activities with Success in Agile Development of Physical Products. In: Elsevier B.V. (ed) Procedia CIRP 109 (2022), pp 146–154. https://doi.org/10.1016/j.procir.2022.05.228

Bertoni A, Larsson T, Larsson J, Elfsberg J (2017) Mining data to design value: A demonstrator in early design. In: Proceedings of the 21st International Conference on Engineering Design (ICED17): Design methods and tools. Curran Associates Inc, Red Hook, NY

Bertoni A, Hallstedt SI, Dasari SK, Andersson P (2020) Integration of value and sustainability assessment in design space exploration by machine learning: An aerospace application. Design Science 6. https://doi.org/10.1017/dsj.2019.29

Boehm BW (1979) Guidelines for Verifying and Validating Software Requirements and Design Specifications. https://ieeexplore.ieee.org/stamp/stamp.jsp?arnumber=1695100. Accessed 16 July 2025

Booth TM, Ghosh S (eds) (2023) Machine learning model cards toward model-based system engineering analysis of resource-limited systems, Proceedings of SPIE - The International Society for Optical Engineering, Vol. 12547. https://doi.org/10.1117/12.2678355

Bröhl AP, Dröschel W (1995) Das V-Modell: Der Standard für die Softwareentwicklung mit Praxisleitfaden. Oldenbourg Wissenschaftsverlag

Buechler T, Kolter M, Hallweger L, Zaeh MF (2022) Predictive cost comparison of manufacturing technologies through analyzing generic features in part screening. CIRP Journal of Manufacturing Science and Technology 38:299–319. https://doi.org/10.1016/j.cirpj.2022.04.012

Calleya J, Pawling R, Ryan C, Gaspar HM (eds) (2016) Using Data Driven Documents (D3) to explore a Whole Ship Model. https://doi.org/10.1109/SYSOSE.2016.7542947

Camuz S, Liljerehn A, Wärmefjord K, Söderberg R (2024) Algorithm for Detecting Load-Carrying Regions Within the Tip Seat of an Indexable Cutting Tool. J. Comput. Inf. Sci. Eng. 24. https://doi.org/10.1115/1.4064255

Chen Z, Zhao Y, Yang J, Wang Y, Dui H (2024) A novel degradation model and reliability evaluation methodology based on two-phase feature extraction: An application to marine lubricating oil pump. Reliability Engineering and System Safety 243. https://doi.org/10.1016/j.ress.2023.109883

Cheung WM, Marsh R, Newnes LB, Mileham AR, Lanham JD (2015) Cost data modelling and searching to support low-volume, high-complexity, long-life defence system development. Proceedings of the Institution of Mechanical Engineers, Part B: Journal of Engineering Manufacture 229:835–846. https://doi.org/10.1177/0954405414534226

Chiu MC, Lin KZ (2018) Utilizing text mining and Kansei Engineering to support data-driven design automation at conceptual design stage. Advanced Engineering Informatics 38:826–839. https://doi.org/10.1016/j.aei.2018.11.002

Choi AJ, Park J, Han J-H (2022) Automated Aerial Docking System Using Onboard Vision-Based Deep Learning. Journal of Aerospace Information Systems 19:421–436. https://doi.org/10.2514/1.I011053

Ciklamini M, Cejnek M (2024) Reinforcement learning inclusion to alter design sequence of finite element modeling. Multiscale and Multidisciplinary Modeling, Experiments and Design. https://doi.org/10.1007/s41939-024-00493-5

Danglade F, Pernot J-P, Véron P (2014) On the use of Machine Learning to Defeature CAD Models for Simulation. Comput.-Aided Des. Appl. 11:358–368. https://doi.org/10.1080/16864360.2013.863510

Ding H, Zhang YT, Kong XN, Shi YJ, Hu ZH, Zhou ZY, Lu R (2022) Data-driven bending fatigue life forecasting and optimization via grinding Top-Rem tool parameters for spiral bevel gears. Advanced Engineering Informatics 53. https://doi.org/10.1016/j.aei.2022.101724

Dizor R, Raj A, Stewart T, Gonzalez B (eds) (2023) Anatomical exoskeleton for movement approximation with neural networks, Proceedings of SPIE - The International Society for Optical Engineering. https://doi.org/10.1117/12.2668383

Douthwaite M, Kelly T (2017) Establishing verification and validation objectives for safety-critical Bayesian networks. In: Proceedings - 2017 IEEE 28th International Symposium on Software Reliability Engineering Workshops, ISSREW, pp 302–309. https://doi.org/10.1109/ISSREW.2017.60

Du X, Jiang B, Zhu F (2021) A new method for vehicle system safety design based on data mining with uncertainty modeling. Eng. Struct. 247. https://doi.org/10.1016/j.engstruct.2021.113184

Dworschak F, Kügler P, Schleich B, Wartzack S (2019) Integrating the mechanical domain into seed approach. In: Proceedings of the 22nd International Conference on Engineering Design (ICED19), pp 2587–2596. https://doi.org/10.1017/dsi.2019.265

Ellis C, Wigness M, Fiondella L (2022) A Mapping of Assurance Techniques for Learning Enabled Autonomous Systems to the Systems Engineering Lifecycle. In: Proceeding - 2022 IEEE International Conference on Assured Autonomy, ICAA, pp 28–35. https://doi.org/10.1109/ICAA52185.2022.00013

Fang C, Zhu Y, Le Fang, Long Y, Lin H, Cong Y, Wang SJ (2025) Generative AI-enhanced human-AI collaborative conceptual design: A systematic literature review. Design Studies 97:101300. https://doi.org/10.1016/j.destud.2025.101300

Fazeli HR, Peng Q (2024) Product Concept Development and Evaluation Using Multiagent Reinforcement Learning. IEEE Transactions on Engineering Management 71:8701–8716. https://doi.org/10.1109/TEM.2024.3399773

Ferreira F, Vieira R, Costa R, Santos R, Moret M, Murari TB (eds) (2021) Data-Driven Hardware-in-the-Loop (HIL) Testing Prioritization. https://doi.org/10.4271/2020-36-0140

Figoli FA, Bruggeman R, Rampino L, Ciuccarelli P (2025) AI-against-design map: A systematic review of 20 years of AI-focused studies in design research. Design Studies 96:101279. https://doi.org/10.1016/j.destud.2024.101279

Fusaro R, Viola N (2018) Preliminary reliability and safety assessment methodology for trans-atmospheric transportation systems. Aircraft Engineering and Aerospace Technology 90:639–651. https://doi.org/10.1108/AEAT-11-2016-0214

Gamage K, Arachchige JK, Siriwardhana S, Kulasekera AL, Dassanayake P (2023) Systems Engineering-Based Approach for the Development of a Strawberry Harvesting Robot. In: 2023 Moratuwa Engineering Research Conference (MERCon), pp 586–591. https://doi.org/10.1109/MERCon60487.2023.10355526

Gao Z, Saxen H, Gao C (2013) Guest Editorial: Special section on data-driven approaches for complex industrial systems. IEEE Transactions on Industrial Informatics 9:2210–2212. https://doi.org/10.1109/TII.2013.2281002

Garriga AG, Mainini L, Ponnusamy SS (2019) A machine learning enabled multi-fidelity platform for the integrated design of aircraft systems. Journal of Mechanical Design 141. https://doi.org/10.1115/1.4044401

Gay A, Iung B, Voisin A, Do P, Bonidal R, Khelassi A (2021) A Short Review on the Integration of Expert Knowledge in Prognostics for PHM in Industrial Applications. In: 5th International Conference on System Reliability and Safety, ICSRS, pp 286–292. https://doi.org/10.1109/ICSRS53853.2021.9660645

Ge H, Bakir AH, Yadav S, Kang Y, Parameswaran S, Zhao P (2021) CFD Optimization of the Pre-Chamber Geometry for a Gasoline Spark Ignition Engine. Frontiers in Mechanical Engineering 6. https://doi.org/10.3389/fmech.2020.599752

Geffner H (2018) Model-free, Model-based, and General Intelligence. Invited talk. IJCAI

Georgiou A, Haritos G, Fowler M, Imani Y (2016a) Advanced phase powertrain design attribute and technology value mapping. Journal of Engineering, Design and Technology 14:115–133. https://doi.org/10.1108/JEDT-05-2014-0031

Georgiou A, Haritos G, Fowler M, Imani Y (2016b) Attribute and technology value mapping for conceptual product design phase. Proc. Inst. Mech. Eng. Part C J. Mech. Eng. Sci. 230:1745–1756. https://doi.org/10.1177/0954406215585595

Gerschütz B, Goetz S, Wartzack S (2023a) AI4PD—Towards a Standardized Interconnection of Artificial Intelligence Methods with Product Development Processes. Applied Sciences (Switzerland) 13. https://doi.org/10.3390/app13053002

Gerschütz B, Sauer C, Kormann A, Nicklas SJ, Goetz S, Roppel M, Tremmel S, Paetzold-Byhain K, Wartzack S (2023b) Digital Engineering Methods in Practical Use during Mechatronic Design Processes. Designs 7. https://doi.org/10.3390/designs7040093

Graening L, Sendhoff B (2014) Shape mining: A holistic data mining approach for engineering design. Advanced Engineering Informatics 28:166–185. https://doi.org/10.1016/j.aei.2014.03.002

Gräßler I, Oleff C, Preuß D (2022) Proactive Management of Requirement Changes in the Development of Complex Technical Systems. Applied Sciences (Switzerland) 12. https://doi.org/10.3390/app12041874

Guariniello C, Marsh TB, Porter R, Crumbly C, Delaurentis DA (eds) (2020) Artificial Intelligence Agents to Support Data Mining for SoS Modeling of Space Systems Design. https://doi.org/10.1109/AERO47225.2020.9172802

Guba EG (1981) Criteria for assessing the trustworthiness of naturalistic inquiries. ECTJ 29:75–91. https://doi.org/10.1007/BF02766777

Guba EG, Lincoln YS (1982) Epistemological and methodological bases of naturalistic inquiry. ECTJ 30:233–252. https://doi.org/10.1007/BF02765185

Gurtner M, Zips P, Heinz T, Atak M, Ophey J, Seiler-Thull D, Kugi A (2021) Efficient oscillation detection for verification of mechatronic closed-loop systems using search-based testing. Mechanical Systems and Signal Processing 163. https://doi.org/10.1016/j.ymssp.2021.108112

Gussen LC, Ellerich M, Schmitt RH (2019) Prediction of perceived quality through the development of a robot-supported multisensory measuring system. In: Procedia CIRP, pp 368–373. https://doi.org/10.1016/j.procir.2019.04.206

Habib MK, Ayankoso SA, Nagata F (2021) Data-Driven Modeling: Concept, Techniques, Challenges and a Case Study. In: 2021 IEEE International Conference on Mechatronics and Automation (ICMA), pp 1000–1007. https://doi.org/10.1109/ICMA52036.2021.9512658

Hajiha M, Liu X, Lee YM, Ramin M (2022) A physics-regularized data-driven approach for health prognostics of complex engineered systems with dependent health states. Reliability Engineering and System Safety 226. https://doi.org/10.1016/j.ress.2022.108677

He C, Li ZK, Wang S, Liu DZ (2021) A systematic data-mining-based methodology for product family design and product configuration. Advanced Engineering Informatics 48. https://doi.org/10.1016/j.aei.2021.101302

Hesse C, Walther J-N, Allebrodt P, Wandel M (eds) (2021) Integration of multi-physics analysis into the cabin design process using virtual reality. https://doi.org/10.2514/6.2021-2776

Hoefer MJ, Frank MC (2018) Automated manufacturing process selection during conceptual design. J Mech Des 140. https://doi.org/10.1115/1.4038686

How DNT, Hannan MA, Hossain Lipu MS, Ker PJ (2019) State of Charge Estimation for Lithium-Ion Batteries Using Model-Based and Data-Driven Methods: A Review. IEEE Access 7:136116–136136. https://doi.org/10.1109/ACCESS.2019.2942213

Hu Z, Du X (2017) System reliability analysis with in-house and outsourced components. In: 2017 2nd International Conference on System Reliability and Safety (ICSRS), pp 146–150. https://doi.org/10.1109/ICSRS.2017.8272811

Huo Y, Liu J, Xiong J, Xiao W, Zhao J (2022) Machine learning and CBR integrated mechanical product design approach. Advanced Engineering Informatics 52. https://doi.org/10.1016/j.aei.2022.101611

Jacobs G, Konrad C, Berroth J, Zerwas T, Höpfner G, Spütz K (2022) Function-Oriented Model-Based Product Development. In: Design Methodology for Future Products. Springer, Cham, pp 243–263. https://doi.org/10.1007/978-3-030-78368-6_13

Janson V, Ahlbrecht A, Durak U (2022) Architectural Challenges in Developing an AI-based Collision Avoidance System. In: 2022 IEEE/AIAA 41st Digital Avionics Systems Conference (DASC), pp 1–8. https://doi.org/10.1109/DASC58513.2023.10311177

Jing LT, Tian CL, He S, Feng D, Jiang SF, Lu CF (2023) Data-driven implicit design preference prediction model for product concept evaluation via BP neural network and EEG. Advanced Engineering Informatics 58. https://doi.org/10.1016/j.aei.2023.102213

Jordan MI, Mitchell TM (2015) Machine learning: Trends, perspectives, and prospects

Jung BC, Park J, Oh H, Kim J, Youn BD (2015) A framework of model validation and virtual product qualification with limited experimental data based on statistical inference. Structural and Multidisciplinary Optimization 51:573–583. https://doi.org/10.1007/s00158-014-1155-2

Kabir S, Walker M, Papadopoulos Y (2018) Dynamic system safety analysis in HiP-HOPS with Petri Nets and Bayesian Networks. Safety Science 105:55–70. https://doi.org/10.1016/j.ssci.2018.02.001

Kim S, Choi J-H, Kim NH (2022) Data-driven prognostics with low-fidelity physical information for digital twin: physics-informed neural network. Structural and Multidisciplinary Optimization 65. https://doi.org/10.1007/s00158-022-03348-0

Kim D, Azad MM, Khalid S, Kim HS (2023) Data-driven surrogate modeling for global sensitivity analysis and the design optimization of medical waste shredding systems. Alexandria Engineering Journal 82:69–81. https://doi.org/10.1016/j.aej.2023.09.077

Knödler J, Könen C, Muhl P, Rudolf T, Sax E, Reuss H-C, Eckstein L, Hohmann S (2023) The Potential of Data-Driven Engineering Models: An Analysis Across Domains in the Automotive Development Process. https://www.sae.org/publications/technical-papers/content/2023-01-0087/. Accessed 16 July 2025

Krahe C, Marinov M, Schmutz T, Hermann Y, Bonny M, May M, Lanza G AI based geometric similarity search supporting component reuse in engineering design. In: Procedia CIRP 2022, pp 275–280. https://doi.org/10.1016/j.procir.2022.05.249

Krahe C, Kalaidov M, Doellken M, Gwosch T, Kuhnle A, Lanza G, Matthiesen S (2020a) AI-Based knowledge extraction for automatic design proposals using design-related patterns. In: Procedia CIRP, pp 397–402. https://doi.org/10.1016/j.procir.2021.05.093

Krahe C, Bräunche A, Jacob A, Stricker N, Lanza G (2020b) Deep Learning for Automated Product Design. In: Procedia CIRP, pp 3–8. https://doi.org/10.1016/j.procir.2020.01.135

Kreis A, Hirz M, Rossbacher P (2020) Cad-automation in automotive development – potentials, limits and challenges. Comput.-Aided Des. Appl. 18:849–863. https://doi.org/10.14733/cadaps.2021.849-863

Kyamakya K, Tavakkoli V, McClatchie S, Arbeiter M, van Scholte Mast BG (2022a) A Comprehensive "Real-World Constraints"-Aware Requirements Engineering Related Assessment and a Critical State-of-the-Art Review of the Monitoring of Humans in Bed. Sensors 22. https://doi.org/10.3390/s22166279

Kyamakya K, Tavakkoli V, McClatchie S, Arbeiter M, van Scholte Mast BG (2022b) A Comprehensive "Real-World Constraints"-Aware Requirements Engineering Related Assessment and a Critical State-of-the-Art Review of the Monitoring of Humans in Bed. Sensors 22. https://doi.org/10.3390/s22166279

Lakoju M, Ajienka N, Khanesar MA, Burnap P, Branson DT (2021) Unsupervised learning for product use activity recognition: An exploratory study of a "chatty device". Sensors 21. https://doi.org/10.3390/s21154991

Lambert NA, Ferrio KB, Goodman DL (2014) A time-domain modeling and simulation framework for comparative analysis of prognostics, reliability and robustness in system design. In: ANNUAL CONFERENCE OF THE PROGNOSTICS AND HEALTH MANAGEMENT SOCIETY, pp 754–761

Lanus E, Hernandez I, Dachowicz A, Freeman LJ, Grande M, Lang A, Panchal JH, Patrick A, Welch S (2021) Test and Evaluation Framework for Multi-Agent Systems of Autonomous Intelligent Agents. In: 16th International System of Systems Engineering Conference, SoSE, pp 203–209. https://doi.org/10.1109/SOSE52739.2021.9497472

Li X, Shao Y, Liu Y (eds) (2015) Takagi-Sugeno model based simulation data mining for efficient product design, 1B-2015. https://doi.org/10.1115/DETC2015-47040

Li Z, Wu G (eds) (2018) A Text Mining based Reliability Analysis Method in Design Failure Mode and Effect Analysis. https://doi.org/10.1109/ICPHM.2018.8448909

Li Y, Roy U, Saltz JS (2019a) Towards an integrated process model for new product development with data-driven features (NPD 3 ). Research in Engineering Design 30:271–289. https://doi.org/10.1007/s00163-019-00308-6

Li W, Liu D (2022) An Approach of Rolling Bearing Remaining Useful Life Prediction based on Adaptive Ensemble Model. In: 2022 International Conference on Sensing, Measurement & Data Analytics in the era of Artificial Intelligence (ICSMD), pp 1–5. https://doi.org/10.1109/ICSMD57530.2022.10058425

Li Q, Wei H, Yu C, Wang S (2022) Data and model-based triple V product development framework and methodology. Enterprise Information Systems 16:1867900. https://doi.org/10.1080/17517575.2020.1867900

Li F, Chen X, Xu P, Fan Z, Wang Q, Lyu C, Zhang Q, Yu H, Wu H (2023) Optimal design of thin-layered composites for type IV vessels: Finite element analysis enhanced by ANN. Thin-Walled Struct 187. https://doi.org/10.1016/j.tws.2023.110752

Li YP, Roy U, Saltz JS (2019b) Towards an integrated process model for new product development with data-driven features (NPD$^3$). Research in Engineering Design 30:271–289. https://doi.org/10.1007/s00163-019-00308-6

Lin K-Z, Chiu M-C (2017) Utilizing text mining and kansei engineering to support data-driven design automation. In: Advances in Transdisciplinary Engineering, pp 949–958. https://doi.org/10.3233/978-1-61499-779-5-949

Lincoln YS, Guba EG (1985) Naturalistic Inquiry, SAGE

Liu Y, Goebel K (eds) (2018) Information fusion for national airspace system prognostics: A NASA ULI project

Liu G, Wang C, Jia Z, Wang K, Ma W, Li Z (2021) A Rapid Design and Fabrication Method for a Capacitive Accelerometer Based on Machine Learning and 3D Printing Techniques. IEEE Sensors Journal 21:17695–17702. https://doi.org/10.1109/JSEN.2021.3085743

Lombardi A, Zampieri L, Agrawal M, Singhal M, Tschammer T von (eds) (2024) Optimization of Power Module Cooling Plate: An Application of Deep Learning for Thermal Management Devices. https://doi.org/10.4271/2024-01-2583

Long HA, French DP, Brooks JM (2020) Optimising the value of the critical appraisal skills programme (CASP) tool for quality appraisal in qualitative evidence synthesis. Research Methods in Medicine & Health Sciences 1:31–42. https://doi.org/10.1177/2632084320947559

Lostado-Lorza R, Escribano-García R, Fernández-Martínez R, Illera-Cueva M, Mac Donald BJ (2014) Combination of the finite element method and data mining techniques to design and optimize bearings. In: Advances in Intelligent Systems and Computing, pp 165–174. https://doi.org/10.1007/978-3-319-07995-0_17

Luo J (2023) Data-Driven Innovation: What Is It? IEEE TRANSACTIONS ON ENGINEERING MANAGEMENT 70:784–790. https://doi.org/10.1109/TEM.2022.3145231

Maslyaev M, Hvatov A, Kalyuzhnaya A (2020) Data-Driven Partial Differential Equations Discovery Approach for the Noised Multi-dimensional Data. In: Springer, Cham, pp 86–100. https://doi.org/10.1007/978-3-030-50417-5_7

Mosallam A, Medjaher K, Zerhouni N (eds) (2015) Component based data-driven prognostics for complex systems: Methodology and applications | IEEE Conference Publication | IEEE Xplore. IEEE, First International Conference on Reliability Systems Engineering (ICRSE). https://doi.org/10.1109/ICRSE.2015.7366504

Mostafapour A, Akbari A, Nakhaei MR (2017) Application of response surface methodology for optimization of pulsating blank holder parameters in deep drawing process of Al 1050 rectangular parts. Int J Adv Manuf Technol 91:731–737. https://doi.org/10.1007/s00170-016-9781-z

Mount NJ, Maier HR, Toth E, Elshorbagy A, Solomatine D, Chang F-J, Abrahart RJ (2016) Data-driven modelling approaches for socio-hydrology: opportunities and challenges within the Panta Rhei Science Plan. Hydrological Sciences Journal:1–17. https://doi.org/10.1080/02626667.2016.1159683

Mun J, J. Lim, Y. Kwak, B. Kang, K. K. Choi, D. -H. Kim (2021) Reliability-Based Design Optimization of a Permanent Magnet Motor Under Manufacturing Tolerance and Temperature Fluctuation. IEEE Transactions on Magnetics 57:1–4. https://doi.org/10.1109/TMAG.2021.3063161

Niu G, Tang S, Liu Z, Zhao G, Zhang B (eds) (2018) Fault diagnosis and prognosis based on deep belief network and particle filtering

Nosck B, Ehlert A, Nayeri CN, Morzynski M (2023) Analysis, Modeling, and Control of the Cylinder Wake. In: Mendez MA, Laniro A, Noack BR, Brunton SL (eds) Data-Driven Fluid Mechanics: Combining First Principles and Machine Learning. Cambridge University Press, pp 3–19. https://doi.org/10.1017/9781108896214.004

Nüßgen A, Degen R, Irmer M, Richter F, Boström C, Ruschitzka M (2024) Leveraging Robust Artificial Intelligence for Mechatronic Product Development : A Literature Review. International Journal of Intelligence Science 14:1–21

Page MJ, McKenzie JE, Bossuyt PM, Boutron I, Hoffmann TC, Mulrow CD, Shamseer L, Tetzlaff JM, Akl EA, Brennan SE, Chou R, Glanville J, Grimshaw JM, Hróbjartsson A, Lalu MM, Li T, Loder EW, Mayo-Wilson E, McDonald S, McGuinness LA, Stewart LA, Thomas J, Tricco AC, Welch VA, Whiting P, Moher D (2021) The PRISMA 2020 statement: an updated guideline for reporting systematic reviews. BMJ 372:n71. https://doi.org/10.1136/bmj.n71

Palm H, Holzmann J (2018) Hyper Space Exploration A Multicriterial Quantitative Trade-Off Analysis for System Design in Complex Environment. In: 2018 IEEE International Systems Engineering Symposium (ISSE), pp 1–6. https://doi.org/10.1109/SysEng.2018.8544435

Papakonstantinou N, Proper S, O'Halloran B, Tumer IY (eds) (2014) Simulation based machine learning for fault detection in complex systems using the Functional Failure Identification and Propagation framework, 1B. https://doi.org/10.1115/DETC201434628

Petersen P, Thorgeirsson AT, Scheubner S, Otten S, Gauterin F, Sax E (2019) Training and validation methodology for range estimation algorithms. In: VEHITS 2019 - Proceedings of the 5th International Conference on Vehicle Technology and Intelligent Transport Systems, pp 434–443. https://doi.org/10.5220/0007717004340443

Pons W, Cordero SS, Vingerhoeds R (2021) Design Structure Matrix Generation from Open-source MBSE Tools. In: 2021 IEEE International Symposium on Systems Engineering (ISSE), pp 1–8. https://doi.org/10.1109/ISSE51541.2021.9582525

Rabe M, Anacker H, Dumitrescu R (eds) (2014) Methodology for the identification of solution patterns within mechatronic systems. https://doi.org/10.1109/REM.2014.6920238

Raoalthi T, Reddy H, Manu K, Manuel N (eds) (2023) Design Optimization in Automotive Product Development Using AI/ML Algorithms. https://doi.org/10.1109/ITEC-India59098.2023.10471413

Rashid M, Anwar MW, Khan AM (2015) Identification of trends for model based development of embedded systems. In: 2015 12th International Symposium on Programming and Systems (ISPS), pp 1–8. https://doi.org/10.1109/ISPS.2015.7245004

Rastogi S, Milind TR, Marsh K (eds) (2023) A Reduced Order Model for Prediction of the Noise Radiated by a High-Speed EV Transmission using Statistical Energy Analysis. https://doi.org/10.4271/2023-01-1113

Raz AK, Guariniello C, Blasch E, Mian Z (2021) An overview of systems engineering challenges for ai-enabled aerospace systems. In: AIAA Scitech 2021 Forum, pp 1–11

Rizzo DB, Blackburn MR (2019) Harnessing Expert Knowledge: Defining Bayesian Network Model Priors From Expert Knowledge Only—Prior Elicitation for the Vibration Qualification Problem. IEEE Systems Journal 13:1895–1905. https://doi.org/10.1109/JSYST.2019.2892942

Rosen KM, Pattipati KR (2023) Operating Digital Twins Within an Enterprise Process. In: The Digital Twin, vol 2, pp 599–659. https://doi.org/10.1007/978-3-031-21343-4_22

Sekar R, Jacome O, Chrstos J, Stockar S (eds) (2022) Assessment of Driving Simulators for Use in Longitudinal Vehicle Dynamics Evaluation. https://doi.org/10.4271/2022-01-0533

Shabestari SS, Herzog M, Bender B (2019) A survey on the applications of machine learning in the early phases of product development. In: pp 2437–2446. https://doi.org/10.1017/dsi.2019.250

Shang L, Jia Y, Liu J (2020) Analysis of mechanical fault diagnosis method according to signal deep autoencoder. International Journal of Mechatronics and Applied Mechanics 1:121–129

Singh AA, Harikrishnan CI, Tiwari SK, Sharma S (2021) A multi-entropy fusion approach for rolling bearing fault diagnosis integrated with variational mode decomposition. In: 2021 8th International Conference on Signal Processing and Integrated Networks (SPIN), pp 367–372. https://doi.org/10.1109/SPIN52536.2021.9566100

Sohier H, Lamothe P, Guermazi S, Yagoubi M, Menegazzi P, Maddaloni A (2021) Improving simulation specification with MBSE for better simulation validation and reuse. Systems Engineering 24:425–438. https://doi.org/10.1002/sys.21594

Son H, Lee G, Kang K, Kang YJ, Youn BD, Lee I, Noh Y (2020) Industrial issues and solutions to statistical model improvement: a case study of an automobile steering column. Structural and Multidisciplinary Optimization 61:1739–1756. https://doi.org/10.1007/s00158-020-02526-2

Spruegel TC, Rothfelder R, Bickel S, Grauf A, Sauer C, Schleich B, Wartzack S (eds) (2018) Methodology for plausibility checking of structural mechanics simulations using Deep Learning on existing simulation data

Su Y, Wang H, Wang J, Liang X (2019) A Three-Dimensional Information Flow Modeling Method for Integrated Fault Diagnosis and Maintenance of Complex System. In: Prognostics and System Health Management Conference (PHM-Paris), pp 51–56. https://doi.org/10.1109/PHM-Paris.2019.00017

Synnes EL, Welo T (2023) Data-driven product optimization capabilities to enhance sustainability and environmental compliance in a marine manufacturing context. Concurrent Engineering Research and Applications 31:113–125. https://doi.org/10.1177/1063293X231217543

Talal M, Ramli KN, Zaidan AA, Zaidan BB, Jumaa F (2020) Review on car-following sensor based and data-generation mapping for safety and traffic management and road map toward ITS. Vehicular Communications 25:100280. https://doi.org/10.1016/j.vehcom.2020.100280

Thukaram P, Mohan S (eds) (2019) Digital Twins for Prognostic Profiling. https://doi.org/10.4271/2019-28-2456

Tong A, Sainsbury P, Craig J (2007) Consolidated criteria for reporting qualitative research (COREQ): a 32-item checklist for interviews and focus groups. Int J Qual Health Care 19:349–357. https://doi.org/10.1093/intqhc/mzm042

Umaras E, Barari A, Tsuzuki M (2021) Tolerance analysis based on Monte Carlo simulation: a case of an automotive water pump design optimization. J Intell Manuf 32:1883–1897. https://doi.org/10.1007/s10845-020-01695-7

Vasantha G, Purves D, Quigley J, Corney J, Sherlock A, Randika G (2021) Common design structures and substitutable feature discovery in CAD databases. Advanced Engineering Informatics 48. https://doi.org/10.1016/j.aei.2021.101261

VDI/VDE 2206 (2021) Entwicklung mechatronischer und cyber-physischer Systeme. VEREIN DEUTSCHER INGENIEURE

Verma A, Singhal N (2024) Integrating Artificial Intelligence for Adaptive Decision-Making in Complex System. In: pp 95–105. https://doi.org/10.1007/978-981-99-9521-9_8

Villarejo R, Johansson C-A, Galar D, Sandborn P, Kumar U (2016) Context-driven decisions for railway maintenance. Proceedings of the Institution of Mechanical Engineers, Part F: Journal of Rail and Rapid Transit 230:1469–1483. https://doi.org/10.1177/0954409715607904

Wagenmann S, Züfle M, Weidinger F, Heinzelmann L, Rapp S, Bursac N, Krause D, Albers A (2023) Data-driven modelling of the functional level in model-based systems engineering - Optimization of module scopes in modular development. In: Procedia CIRP 119, pp 1047–1052. https://doi.org/10.1016/j.procir.2023.03.147

Wang CC, Ye C, Bi YR, Wang JX, Han YW (2023) Application of mechanical product design parameter optimization based on machine learning in identification. Prod Plann Control. https://doi.org/10.1080/09537287.2022.2160388

Wöhr F, Uhri E, Königs S, Trauer J, Stanglmeier M, Zimmermann M (2023) Coordination and complexity: an experiment on the effect of integration and verification in distributed design processes. Design Science 9. https://doi.org/10.1017/dsj.2022.26

Wu J (2024) An exploratory study of v-model in building ml-enabled software: A systems engineering perspective. In: pp 30–40. https://doi.org/10.1145/3644815.3644951

Wynn DC, Clarkson PJ (2018) Process models in design and development. Res Eng Design 29:161–202. https://doi.org/10.1007/s00163-017-0262-7

Yang B, Xiao R-B (2021) Data-Driven Product Design and Axiomatic Design. In: 2021 IEEE International Conference on Progress in Informatics and Computing (PIC), pp 489–493. https://doi.org/10.1109/PIC53636.2021.9687021

Yuan C, Marion T, Moghaddam M (2022) Leveraging End-User Data for Enhanced Design Concept Evaluation: A Multimodal Deep Regression Model. J Mech Des 144. https://doi.org/10.1115/1.4052366

Zhang K-K, K. -M. Hu, L. -C. Yin, D. -M. Yan, B. Wang (2015) CAD Parts-Based Assembly Modeling by Probabilistic Reasoning. In: 2015 14th International Conference on Computer-Aided Design and Computer Graphics (CAD/Graphics), pp 89–96. https://doi.org/10.1109/CADGRAPHICS.2015.29

Zhang C, Zhou G (2019) A view-based 3D CAD model reuse framework enabling product lifecycle reuse. Adv Eng Software 127:82–89. https://doi.org/10.1016/j.advengsoft.2018.09.001

Zhang C, Kwon YP, Kramer J, Kim E, Agogino AM (2017) Concept Clustering in Design Teams: A Comparison of Human and Machine Clustering. Journal of Mechanical Design 139. https://doi.org/10.1115/1.4037478

Zhang C, Zhou G, Yang H, Xiao Z, Yang X (2020) View-Based 3-D CAD Model Retrieval with Deep Residual Networks. IEEE Trans. Ind. Inf. 16:2335–2345. https://doi.org/10.1109/TII.2019.2943195

Zhang J, Zhang K, An Y, Luo H, Yin S (2024) An Integrated Multitasking Intelligent Bearing Fault Diagnosis Scheme Based on Representation Learning Under Imbalanced Sample Condition. IEEE Transactions on Neural Networks and Learning Systems 35:6231–6242. https://doi.org/10.1109/TNNLS.2022.3232147

Zhang SW, Cheng Q, Wang CJ, Fang XF, Cheng DJ (2021) A technology of efficient configuration of a ship's critical process equipment. Int J Computer Integr Manuf 34:1370–1381. https://doi.org/10.1080/0951192X.2021.1972462

Zheng H, Feng Y, Gao Y, Tan J, Zheng H, Feng Y, Gao Y, Tan J (2018) A robust predicted performance analysis approach for data-driven product development in the industrial internet of things // A Robust Predicted Performance Analysis Approach for Data-Driven Product Development in the Industrial Internet of Things. Sensors (Switzerland) 18. https://doi.org/10.3390/s18092871

Zheng X, Zhuang C, Xiao S, Qiu Y, Zhang J, Li M (2023) Signal Estimation for Vehicle Body Accelerations Using Piecewise Linear System Identification in the Frequency Domain. Journal of Computing and Information Science in Engineering 23. https://doi.org/10.1115/1.4054306

Zhou Y, Yang L, Li Y, Lu W (2019) Exploring the Data-Driven Modeling Methods for Electrochemical Migration Failure of Printed Circuit Board. In: Proceedings - 2019 Prognostics and System Health Management Conference, PHM-Paris, pp 100–105. https://doi.org/10.1109/PHM-Paris.2019.00025

# Appendix A

Table A1 Overview of the results of the common articles between stages

| No | Reference | Application | Data-Driven Method | Algorithm(s) | Data Type | Stages |
|---|---|---|---|---|---|---|
| **1** | Krahe et al. (2020b) | Identify design patterns specific to a product family by generating point cloud-based concepts from design requirements | Deep learning | Autoencoder network, Generative Adversarial Networks | Numerical data (Point clouds) | **System Design – System Implementation** |
| **2** | Krahe et al. | Geometric similarity search for component reuse | Deep learning | Convolutional neural network and Autoencoder network | Numerical data (CAD) | **System Design – System Implementation** |
| **3** | Lin and Chiu (2017) | Text mining of customer reviews combined with Kansei Engineering for user-driven design. (Conceptual design automation based on customer reviews) | Machine learning | SPSS Modeler text mining algorithm | Textual data (Customer reviews) | **System Design – System Implementation** |
| **4** | Bertoni et al. (2020) | Design space exploration of Aerospace engine component (Turbine Rear Structure) design considering value and sustainability | Machine learning and Statistical | Random Forest (RF), regression analysis, DOE, surrogate models | Various (CAE Simulation Data Material/Cost Databases, ICAO Emission Data, and Maintenance Cost Data) | **System Design – System Integration** |
| **5** | Booth and Ghosh (2023) | Support of system analysis, integration and design for resource-limited systems using ML models | Deep learning and Machine learning | Recurrent neural network (RNN), Monte Carlo analysis (Model Cards) | Numerical data (Sensor fusion data) and Textual (model documentation) | **System Design – System Integration** |
| **6** | Akkaya et al. (2016) | Segregation of domains of expertise and cross-cutting concerns in systems engineering<br><br>(Disaster response robotics, industrial CPS modeling, and swarm intelligence.) | Machine Learning | Actor-oriented and Aspect-oriented modeling techniques. | Numerical data (Simulation) | **System Design – System Integration** |
| **7** | Lambert et al. (2014) | Comparative analysis of prognostics, reliability, and robustness /or (Reliability and robustness assessment, trade-space analysis for PHM of dynamic systems) | Statistical | Monte Carlo and Life State Approach approaches | Numerical data (Simulated time-series) | **System Design - System Integration** |
| **8** | Mun et al. (2021) | Electromagnetic System Optimization under Uncertainty (Monte Carlo and surrogate modeling for robust rotor design under uncertainties.) | Machine Learning and Statistical | RBDO method, Dynamic Kriging (DKG) Surrogate Modeling+ Monte Carlo Simulation | Numerical data (design parameters and simulation data from MagNet FEA Software) | **System Implementation - System Integration** |
| **9** | (Lanus et al. 2021) | Hierarchical testing framework for integrating local and global testing methods for scalability and adaptability of (multi-agent systems) satellite constellations and dynamic operational environments. . | Statistical | Experiments (DOE) and Combinatorial Interaction Testing (CIT).<br>Optimal Learning | Numerical data (Simulated data of satellite network MAS) | **System Integration - Validation** |
| **10** | (Ferreira et al. 2021) | Improving test efficiency and decision-making in automotive ECU development. | Statistical/Probabilistic and Machine learning | ETL pipeline, Natural Language Processing (NLP), Data Mining, Knowledge Discovery in Text (KDT), Test Prioritization Algorithms, and HIL Testing Frameworks | Numerical data (Simulation data, Test case performance metrics; defect reports, vehicle requirements, and test artifacts stored across multiple systems). | **System Integration - Validation** |

Table A2 Overview of the results of system design stage

| No. | Reference | Application | DDM | Algorithm(s) | Data Type |
|---|---|---|---|---|---|
| **1** | Akay et al. (2021) | Automated extraction of design requirements | Deep learning | Bidirectional Encoder Representation from Transformers (BERT) | Textual data |
| **2** | Allagui et al. (2023) | Supporting decision making regarding assembly design | Deep learning | Neural network | Numerical data |
| **3** | Yang and Xiao (2021) | N/A | General DDM | N/A | N/A |
| **4** | Bachmayer et al. (2020) | Design and optimization of an underwater robot with regards to efficiency, reliability, and self-diagnostic capabilities | Machine learning | N/A | Numerical data |
| **5** | Batliner et al. (2022) | Analysis of students' design processes and their success | Statistical approach | N/A | Numerical data |
| **6** | Bertoni et al. (2017) | Decision support in early design stages to reduce uncertainty | Machine learning | Multiple linear regression, k-nearest neighbor (KNN), random forest (RF) | Numerical data |
| **7** | Buechler et al. (2022) | Cost prediction based on previous projects' data | Machine learning and Deep Learning | Linear regression, random forest (RF), artificial neural network (ANN) | Numerical data |
| **8** | Calleya et al. (2016) | Prediction of performance | General DDM | N/A | Various |
| **9** | Cheung et al. (2015) | Cost prediction for low-volume, high-complexity, and long-life products | General DDM | N/A | Textual, image and numerical data |
| **10** | Chiu and Lin (2018) | Forecasting customer preferred product designs for conceptual design | Machine learning | Text mining | Textual data |
| **11** | Du et al. (2021) | Creation and evaluation of designs with regards to safety | Machine learning | Decision tree | Textual and numerical data |
| **12** | Dworschak et al. (2019) | Extraction and storage of data from product designs | Deep learning | Artificial neural network (ANN) | Textual and numerical data |
| **13** | Fusaro and Viola (2018) | Safety and reliability assessment during conceptual and preliminary design | Statistical approach | N/A | Numerical data |
| **14** | Garriga et al. (2019) | Design space exploration | Machine learning | K-means | Numerical data |
| **15** | Ge et al. (2021) | CFD optimization | Machine learning | Linear regression models, regression trees, support vector machines (SVM), gaussian process regression | Numerical data |
| **16** | Georgiou et al. (2016a) | Evaluation and objective selection of conceptual designs | General DDM | N/A | Numerical data |
| **17** | Georgiou et al. (2016b) | Evaluation and objective selection of conceptual designs | General DDM | N/A | Numerical data |
| **18** | Gerschütz et al. (2023b)(Review) | N/A | N/A | N/A | N/A |
| **19** | Graening and Sendhoff (2014) | Comparison and analysis of multiple design concepts | Machine learning | N/A | Numerical data |
| **20** | Gräßler et al. (2022) | Proactive management of requirement changes | Deep learning | BERT (Bidirectional Encoder Representations from Transformers) | Textual data |
| **21** | Guariniello et al. (2020) | AI supported document search for systems engineering | Machine learning | Artificial intelligence agents | Textual data |
| **22** | Gurtner et al. (2021) | Analysis of closed-loop mechatronic systems regarding oscillations | Statistical Approach | Bayesian Optimization | Numerical data |
| **23** | Gussen et al. (2019) | Prediction of perceived quality of a product | Machine learning | N/A | Numerical data |
| **24** | Palm and Holzmann (2018) | Design space exploration | Statistical Approach | N/A | Numerical data |
| **25** | Fazeli and Peng (2024) | Design space exploration within concept development | Machine learning | Reinforcement learning | Numerical data |
| **26** | He et al. (2021) | Product family design taking customer requirements into account | Machine learning | Rule based classifier, data mining | Textual data |
| **27** | Huo et al. (2022) | Automatic design based on customer requirements, product information, and design knowledge | Machine learning | K-Nearest-Neighbor (KNN), linear regression, genetic algorithm | Numerical data |
| **28** | Aphirakmethawong et al. (2022)(Review) | N/A | N/A | N/A | N/A |
| **29** | Jing et al. (2023) | Design preference evaluation for concept evaluation | Deep learning | Neural network | Textual data |
| **30** | Kabir et al. (2018) | Model-based safety analysis | Statistical approach | Pandora, Bayesian network, petri net | Numerical data |
| **31** | Kim et al. (2023) | Surrogate modeling for sensitivity analysis and design optimization | Statistical approach | Latin hypercube sampling | Numerical data |
| **32** | Kreis et al. (2020) | CAD automation | Deep learning | Neural network | Numerical data |
| **33** | Lakoju et al. (2021) | Gaining insights into the use of the product by the customer | Machine learning | N/A | Numerical data |
| **34** | Li et al. (2019a) | Effective collaboration in mixed teams | Machine learning | Data mining | N/A |
| **35** | Liu et al. (2021) | Predict the performance from given design parameters and retrieve the design parameters from the customer requirements | Machine learning | Artificial neural network (ANN) | Numerical data |
| **36** | Habib et al. (2021) | Analysis of physical systems | General DDM | N/A | Numerical data |
| **37** | Papakonstantinou et al. (2014) | Fault detection in complex systems | Machine learning | Artificial neural network, decision tree | Numerical data |
| **38** | Rabe et al. (2014) | Identification of solution patterns within mechatronic systems | General DDM | N/A | Various |
| **39** | Umaras et al. (2021) | Tolerance analysis | Statistical approach | Monte Carlo simulation | Numerical data |
| **40** | Pons et al. (2021) | Design Structure Matrix generation from MBSE tools | General DDM | N/A | Textual data |
| **41** | Wagenmann et al. (2023) | Including machine usage data into modelling the functional level of technical systems | General DDM | N/A | Numerical data |
| **42** | Yuan et al. (2022) | Concept evaluation based on large-scale user reviews | Deep learning | Multimodal deep regression | Textual data |
| **43** | Hu and Du (2017) | System reliability analysis for systems outsources and in-house components | Statistical approach | Monte Carlo simulation | Numerical data |
| **44** | Zhang et al. (2017) | Automatic concept clustering | Machine learning | N/A | Textual data |
| **45** | Zheng et al. (2018) | Predicted performance analysis | Statistical approach | Least squares support vector regression | Numerical data |

Table A3 Overview of the results of system implementation stage

| No. | Reference | Application | DDM | Algorithm(s) | Data Type |
|---|---|---|---|---|---|
| **1** | Aschenbrenner and Wartzack (2017) | Statistical tolerance analysis of bearing seats to improve operating clearance and fatigue life. | Statistical analysis | Spearman Correlation | Numerical data (Geometric features) |
| **2** | Askhatova et al. (2023) | Replacing FEM with CNN-based topology optimization to save computational time. | Deep learning | Deep Convolutional Neural Network | Numerical data (CAD) |
| **3** | Camuz et al. (2024) | Detecting load-carrying zones via contact index to improve interface design in early stages. | Statistical analysis | Monte Carlo simulation | Numerical data (CAD) |
| **4** | Ciklamini and Cejnek (2024) | Accelerating FEM-based optimization using reinforcement learning for efficient convergence. | Machine learning | Epsilon-greedy algorithm | Numerical data (CAD features and Simulation data) |
| **5** | Danglade et al. (2014) | ML framework to identify and select features suitable | Deep learning | Neural network | Numerical data (CAD) |
| | | for defeaturing in simulation preparation. | | | |
| **6** | Ding et al. (2022) | Forecasting and optimizing gear fatigue life using DDMs and tool parameters. | General DDM | N/A | Numerical data (CAD) |
| **7** | Hesse et al. (2021) | VR-enhanced design workflow with multi-physics simulation for aircraft cabin systems. | General DDM | N/A | Numerical (Simulation) data |
| **8** | Hoefer and Frank (2018) | Analyzing CAD geometry to automatically suggest suitable manufacturing processes. | Machine learning | Random forest | Numerical data (CAD) |
| **9** | Kim et al. (2023) | Using surrogate models for faster design and optimization of shredder blades. | Machine learning | Latin hypercube sampling | Numerical data (CAD) |
| **10** | Krahe et al. (2020a) | Extracting design logic from CAD model trees to suggest design next steps. | Deep learning | Recurrent Neural Network (RNN) | Numerical data (CAD) |
| **11** | Li et al. (2023) | Coupling FEM with ANN to optimize composite vessel design. | Deep learning | Neural network | Numerical data (Geometric and Manufacturing features) |
| **12** | Li et al. (2015) | Leveraging simulation data with Takagi-Sugeno models to identify key design-performance relationships. | Machine learning | Takagi-Sugeno Model | Numerical data (Simulation) |
| **13** | Lombardi et al. (2024) | Optimizing thermal device geometry using DL to | Deep learning | 3D Deep Learning and Geodesic Convolutional Neural Networks | Numerical data (CFD Simulation) |
| | | minimize pressure loss and maximize cooling. | | | |
| **14** | Lostado-Lorza et al. (2014) | FEM + regression + genetic algorithms to improve bearing design and performance. | Machine learning and Deep learning | Linear Regression, Quadratic Regression, Isotonic Regression, Gaussian Processes, Artificial Neural Networks (MLP), Support Vector Machines (SVM), Regression Trees | Numerical data (FEM) |
| **15** | Mostafapour et al. (2017) | Optimization of deep drawing process using Response Surface Methodology for better formability. | Statistical analysis | Response Surface Methodology (RSM) | Numerical data (Experimental) |
| **16** | Raoalthi et al. (2023) | Integrating AI/ML (ANN, Gaussian regression) into CAD-CAE to optimize automotive products. | Deep learning | Artificial Neural Networks (ANN) | Numerical data (Physical testing and simulation) |
| **17** | Rastogi et al. (2023) | Statistical energy analysis for fast and accurate prediction of acoustic energy flow. | Statistical analysis | Statistical energy analysis | Numerical data (CAD and time series) |
| **18** | Spruegel et al. (2018) | CNNs used to assess plausibility of FEM results via spherical projections. | Deep learning | Convolution neural network (CNN) | Numerical data (FEM simulation) |
| **19** | Thukaram and Mohan (2019) | ML-based anomaly detection using field data to enhance system design via digital twins. | Machine learning | Classification algorithm & Vector autoregression | Numerical data (time series) |
| **20** | Umaras et al. (2021) | Monte Carlo simulations used to evaluate feasibility of geometry changes for cost-saving. | Statistical analysis | Monte Carlo simulation | Numerical data (Geometric features) |
| **21** | Vasantha et al. (2021) | Identifying compatible parts using pattern recognition (e.g., hole patterns) in CAD geometry. | Machine learning | Kullback-Leiber divergence | Numerical data (CAD) |
| **22** | Wang et al. (2023) | ML + NSGA-II used to reduce cavitation in slurry blades via design parameter tuning. | Machine learning | NSGA-II | Numerical data (CFD Simulation) |
| **23** | Zhang and Zhou (2019) | Using ResNet architectures for CAD model retrieval from sketches or views. | Deep learning | ResNet 14,32,50,80, VSC_WCO, ADM_GHD | Numerical data (CAD) and Images (CAD Views) |
| **24** | Zhang et al. (2020) | FilterNet and RankNet used for image-based CAD model matching. | Deep learning | FilterNet, RankNet | Numerical data (CAD) |
| **25** | Zhang et al. (2015) | AI-assisted part recommendation and assembly automation in CAD systems. | Machine learning | Bayesian network | Numerical data (CAD) |
| **26** | Zhang et al. (2021) | Component selection via angle cosine algorithm to match functional requirements. | Machine learning | Angle cosine algorithm | Numerical data (CAD) |

Table A4 Overview of the results of system integration stage

| **No.** | **Reference** | **Application** | **DDM** | **Algorithm(s)** | **Data Type** |
|---|---|---|---|---|---|
| **1** | (Singh et al. 2021) | Fault diagnostic | Statistical and Machine Learning | VMD, MEFA, KNN, RF, Random Forest | Numerical data (time-series vibration signals) |
| **2** | (Ahmed and Chen 2023) | Parameter optimization steam ejector performance for HVAC and refrigeration systems | Deep Learning | Artificial Neural Network (ANN) with optimizers (ADAM, Ftrl, Adadelta), Multi-objective Optimization | Numerical data (Simulated time-series of steam ejector performance) |
| **3** | (Choi et al. 2022) | identify a physical docking system based on a video stream and navigate a aircraft to the docking system | Deep Learning | YOLOv4 | Imaging |
| **4** | (Dizor et al. 2023) | Exoskeleton control for movement approximation. | Machine Learning and Deep Learning | Reinforcement Learning with Neural Networks (Deep Imitation Learning). | Numerical (time-series of EMG and IMU signals). |
| **5** | (Gay et al. 2021) | Integration of expert knowledge to improve prognostics for industrial systems, including a high-pressure water circuit in steel production. | General DDM | N/A | N/A |
| **6** | (Hajiha et al. 2022) | Estimation the health state of a system by modeling it with a physical–statistical model that can be used to extract (hidden) health states | Statistical | Not clear | Numerical (time-series) data |
| **7** | (Gamage et al. 2023) | Image filtering and classification for Strawberry harvesting | Deep Learning | YOLOv4 | Imaging |
| **8** | (Gamage et al. 2023) | Fault Diagnosis | Statistical | Not clear | Numerical (time-series) data |
| **9** | (Kim et al. 2022) | Prognostics with low-fidelity physical information | Deep Learning | Physics Informed Neural Networks (PINNs) | Numerical (time-series) data |
| **10** | (Niu et al. 2018) | Estimation of Capacity | Deep Learning | Deep Belief Network (DBN) and<br>Bayesian estimation | Numerical (time-series) data |
| **11** | (Raz et al. 2021) | AI-Enabled Aerospace Systems | General DDM | N/A | N/A |
| **12** | (Rosen and Pattipati 2023) | Aerospace product development (e.g., S-92 Helicopter, UAVs).<br>And applicable to broader CPS systems | Deep Learning and Statistical | Not algorithm-specific; uses deep learning, CFD modeling, multi-functional causal models, SHM models | Numerical data (sensor and simulation) |
| **13** | (Shang et al. 2020) | Predictive maintenance and fault detection for mechanical systems | Deep Learning | Deep Autoencoder, Sparse Autoencoder (SAE), Compression Autoencoder (CAE), Softmax Classifier | Numerical data (time-series of vibration signal) |
| **14** | (Su et al. 2019) | Fault diagnosis based on human text input | Deep Learning | Ontology Bayesian Network | Textual data |
| **15** | (Synnes and Welo 2023) | Marine Manufacturing (low-volume, high-variance, engineer-to-order production) | General DDM | Not clear | Numerical data (Product data; CAD, PLM, material data) |
| **16** | (Janson et al. 2022) | Aviation collision avoidance in Advanced Air Mobility (AAM) systems | Deep Learning | Feed-Forward Neural Networks (Vertical CAS, Horizontal CAS) | Numerical (multiple independent networks for vertical and horizontal decision planes) |
| **17** | (Li and Liu 2022) | Predictive maintenance of rotating machinery (industrial bearings)- degradation tracking | Machine Learning | Adaptive Ensemble Model (AEM); Dynamic Ensemble Model (DEM) combining Linear, Polynomial, and Dual-Exponential Models | Numerical data (time-series vibration signals |
| **18** | (Wöhr et al. 2023) | | Statistical | Skillings-Mack, Mann–Whitney U, Shapiro–Wilk Tests | Numerical data (Human experimental data; tasks, task completion times, coupling matrices, design iterations). |
| **19** | (Zhou et al. 2019) | Failure prediction in high-density electronic assemblies | Machine Learning | Support Vector Regression (SVR), Gradient Boosting Regression Trees (GBRT), Random Forests (RF) | Numerical data (Time-series of vibration and electrical resistance degradation signals (SIR curves)) |

Table A5 Overview of the results of validation stage

| No. | Reference | Application | DDM | Algorithm(s) | Data Type |
|---|---|---|---|---|---|
| **1** | Bach et al. (2017a) | Automotive: Predictive Cruise Control (PCC) systems to optimize V&V for geolocation-specific scenarios. | Machine-Learning | Classification-tree approach for specification-based selection | Numerical data |
| **2** | Bach et al. (2017b) | Automotive Systems Engineering development process | Statistical | Statistical analysis, Data visualization, Recorded scenario analysis | Numerical data |
| **3** | Chen et al. (2024) | Machinery: Reliability evaluation and remaining useful life prediction for oil pumps | Statistical | Wiener process, Box-Cox transformation, degradation feature extraction, approximate entropy | Numerical data (experimental) |
| **4** | Douthwaite and Kelly (2017) | Safety-critical Bayesian Networks for autonomous systems | Statistical and Machine Learning | Bayesian Networks, Reference model for BN-based systems | Numerical data |
| **5** | Ellis et al. (2022) | Autonomous systems (special focus on automotive) | Statistical, Machine-Learning, Hybrid | Markov Decision Process (MDP) | Numerical and textual data |
| **6** | Jung et al. (2015) | Virtual product qualification with limited experimental data | Statistical | Statistical inference, Hypothesis testing, Maximum Likelihood Estimation (MLE) | Numerical data (experimental) |
| **7** | Kyamakya et al. (2022b) | Monitoring of humans in bed for anomaly detection | Statistical and Deep Learning | Hidden Markov Models, Graph Networks + Deep Learning | Numerical and Textual data |
| **8** | Li and Wu (2018) | Reliability analysis in Design FMEA | Statistical and Deep Learning | Text Mining (with PMI and NPMI) and Neural Network | Textual data |
| **9** | Liu and Goebel (2018) | Prognostics for National Airspace System | Statistical and Deep Learning | Bayesian Entropy Network (BEN), Physics-based Learning, Deep Residual RNN | Numerical data (simulation) |
| **10** | Petersen et al. (2019) | Automotive: Electric vehicle range estimation, addressing driver-specific and driver-unspecific performance for energy consumption prediction | Machine-Learning | k-fold cross validation | Numerical data (experimental) |
| **11** | Sekar et al. (2022) | Driving simulator validation for longitudinal vehicle dynamics | Statistical | Bayesian Networks, Reference model for BN-based systems | Numerical |
| **12** | Sohier et al. (2021) | Automotive: Design and validation of autonomous car | Machine-Learning | Information Theoretic Metric Learning (ITML) | Numerical data (experimental) |
| **13** | Son et al. (2020) | Automotive: Steering columns, specifically addressing vibrational behavior and improving the predictive accuracy of natural frequencies | Machine Learning and Statistical | Kriging-based surrogate modeling for reducing computational costs, statistical model improvement | Numerical data (simulation) |
| **14** | Zheng et al. (2023) | Automotive: Vehicle body durability design to estimate hard-to-measure accelerations and reduce the number of required accelerometers | Statistical | Frequency-domain system identification using noncausal finite impulse response (FIR) | Numerical data (experimental) |

8

9